\documentclass[12pt]{article}

\usepackage{amsmath,amssymb} 
\usepackage{bm}
\usepackage{bbm}
\usepackage{color} %
\usepackage{graphicx}
\usepackage{tabularx}

\usepackage{epstopdf}
\newenvironment{noteblue}{\color{blue}} {\ignorespacesafterend}

\newcommand{\Sec}[1]{Sec.~\ref{#1}}  %

\newcommand{\fig}[1]{Fig.~\ref{#1}}

\newcommand{\tab}[1]{Tab.~\ref{#1}}

\newcommand{\eq}[1]{Eq.~(\ref{#1})}

\begin{document}
\title{
\vspace{-1.0truecm}
\Large\bf
Upper limits on DM annihilation cross sections from the first AMS-02 antiproton data
}
\author{
Hong-Bo Jin$^{a,b}$, %
Yue-Liang Wu$^{a,c,d}$, \ %
and  Yu-Feng Zhou$^{a,c}$
\footnote{Emails: hbjin@bao.ac.cn,  ylwu@itp.ac.cn, yfzhou@itp.ac.cn}
\\ \\
\it $^{a)}$ State Key Laboratory of Theoretical Physics,
\it $^{b)}$ National   \\ 
\it Astronomical  Observatories, 
\it Chinese Academy of Sciences, \\
\it $^{c)}$ Kavli Institute for Theoretical Physics China,
\it Institute   of  \\
\it Theoretical Physics Chinese Academy of Sciences,\\
\it $^{d)}$University of Chinese Academy of Sciences,
\\ 
\it Beijing, 100190, P.R. China
}

\maketitle

\vspace{-1.0truecm}
\begin{abstract}
The first measurement on 
the antiproton to proton ratio by
the AMS-02 collaboration agrees  with 
the expectation from conventional cosmic-ray secondaries
in the kinetic energy range  $\sim 20-100$ GeV,
which can be turned into stringent upper limits on 
the annihilation cross sections of dark matter (DM)  above $\sim 300$ GeV.
Using the GALPROP code, 
we derive the upper limits in various propagation models 
and DM profiles.
We show that 
in the ``conventional'' propagation model, for the  $q\bar q$, $b\bar b$, and 
$W^{+}W^{-}$ final states, the constraints could be more stringent than that derived from
the recent Ferm-LAT gamma-ray data on 
the dwarf spheroidal satellite galaxies.
Making use of  the typical 
minimal, median and maximal models obtained from 
a previous GALPROP based global fit to 
the preliminary AMS-02 data, 
we show that the variation of  the upper limits is about a factor of five.
The possibility of DM contributions to the low and very high energy $\bar p/p$ data 
is discussed.
\end{abstract}

\newpage

Dark matter (DM) is known to contribute to $26.8\%$ of 
the total energy density  of the Universe
\cite{Planck:2015xua}.%
However, its particle nature  remains largely unknown.
If  the DM particles can annihilate or decay into 
the standard model (SM) final states, 
they will contribute to new primary sources of cosmic-ray particles,
and  result in significantly changes in  
the spectra of the  cosmic-ray antiparticles such as positrons and antiprotons,
as  these species  are assumed to be secondaries in
the conventional cosmic-ray propagation theory.
Antiprotons are highly expected from DM annihilation in many DM models,
and  unlikely to be generated from the nearby pulsars.
Compared with cosmic-ray electrons/positrons, 
the cosmic-ray antiprotons loss much less energy during propagation, 
and can travel through longer distance in the Galaxy, 
which makes the antiproton flux more sensitive to the uncertainties in 
the propagation parameters and the DM profiles.

Recently, the AMS-02 collaboration has released 
the first preliminary result of  the cosmic-ray antiproton to proton flux ratio $\bar p /p$
\cite{Ting:AMS}.
The measured  kinetic energies of the antiprotons have been extended to  
$\sim 450$ GeV. 
Although 
the spectrum of $\bar p /p$ at high energies above 100 GeV tend to be relatively flat,
within uncertainties 
the AMS-02 data are consistent with 
the  background of secondary antiprotons,
which can be used to set stringent upper limits on 
the  dark matter (DM) annihilation cross sections,
especially for high mass DM particles.
The constraints on the DM properties from antiprotons have been
investigated previously before  AMS-02
( see eg.
\cite{Hooper:2014ysa,%
Kappl:2014hha,%
Fornengo:2013xda,%
Cirelli:2013hv,%
Jin:2012jn%
} ).
In this work, 
we explore the significance of the new AMS-02 $\bar p/p$ data on constraining the 
annihilation cross sections of the DM particles in 
various propagation models and DM profiles. 
Four representative background models are considered with 
four different DM profiles.
We derive the upper limits using the GALPROP code
and 
show that   
in the ``conventional '' propagation model with  Einasto DM profile, 
the constraints could be more stringent than that derived from
the Ferm-LAT gamma-ray data on 
the dwarf spheroidal satellite galaxies.
Making use of  the typical 
minimal, median and maximal models obtained from 
a previous GALPROP based global fit to the preliminary AMS-02 data,
we show that
the uncertainties on the upper limits is around a factor of five.

We start with a briefly overview on the main features of 
the cosmic-ray propagation within the Galaxy.
The Galactic halo within which  the diffusion processes occur is parametrized by 
a cylinder with radius $R_{h} = 20$ kpc and half-height $Z_{h}=1-20$ kpc.   
The diffusion  equation for the cosmic-ray charged particles reads
(see e.g. \cite{Ginzburg:1990sk})
\begin{align}\label{eq:propagation}
  \frac{\partial \psi}{\partial t} =&
  \nabla (D_{xx}\nabla \psi -\boldsymbol{V}_{c} \psi)
  +\frac{\partial}{\partial p}p^{2} D_{pp}\frac{\partial}{\partial p} \frac{1}{p^{2}}\psi
  -\frac{\partial}{\partial p} \left[ \dot{p} \psi -\frac{p}{3}(\nabla\cdot \boldsymbol{V}_{c})\psi \right]
  \nonumber \\
  & -\frac{1}{\tau_{f}}\psi
  -\frac{1}{\tau_{r}}\psi
  +q(\boldsymbol{r},p)  ,
\end{align}
where $\psi(\boldsymbol{r},p,t)$ is  
the number density per unit of total particle momentum.
For steady-state diffusion, it is assumed that  $\partial  \psi/\partial t=0$.
The number densities of cosmic-ray particles are vanishing at the boundary of the halo,
i.e.,
$\psi(R_{h},z,p)=\psi(R, \pm Z_{h},p)=0$.
The energy dependent spatial diffusion coefficient $D_{xx}$  is parametrized as
$D_{xx}=\beta D_{0} \left(\rho/\rho_{0}\right)^{\delta}$ ,
where $\rho=p/(Ze)$ is the rigidity of the cosmic-ray particle with electric charge $Ze$.
The  power spectral index $\delta$ can have  different values 
$\delta=\delta_{1(2)}$  
when $\rho$ is below (above) a reference rigidity $\rho_{0}$.  
The coefficient  $D_{0}$ is a normalization constant, 
and $\beta=v/c$ is the velocity of the cosmic-ray particle.
The convection term in the diffusion equation is related to 
the drift of cosmic-ray particles from 
the Galactic disc due to the Galactic wind.  
The  diffusion in momentum space is described by 
the reacceleration parameter $D_{pp}$ 
which is related to the  velocity of disturbances in the hydrodynamical plasma, 
and described by the Alfv$\grave{\mbox{e}}$n speed $V_{a}$
\cite{Ginzburg:1990sk}.
In \eq{eq:propagation},
the momentum loss rate is denoted  by $\dot{p}$ which 
could  be due to 
ionization in the interstellar medium neutral matter,
Coulomb scattering off thermal  electrons in ionized plasma,
bremsstrahlung,
synchrotron radiation,
and 
inverse Compton scattering, etc..
The parameter $\tau_{f}(\tau_{r})$ is 
the time scale for fragmentation (radioactive decay) of 
the cosmic-ray nuclei as they interact with interstellar hydrogen and helium.

The spectrum of a primary source term for a cosmic-ray nucleus  $A$  is 
assumed to have a broken power low behaviour 
$dq_{A}(p)/dp \propto
\left( \rho/\rho_{As}\right)^{\gamma_{A}}$
with $\gamma_{A}=\gamma_{A1}(\gamma_{A2})$ for 
the nucleus rigidity $\rho$ below (above) a reference rigidity $\rho_{As}$.
The spatial distribution of the primary sources is taken  from
 Ref.~\cite{Strong:1998pw}.
Secondary antiprotons are created dominantly from 
inelastic $pp$- and $p$A-collisions with the interstellar  gas.
The corresponding source term  reads
\begin{align}
q(p)=
\beta c n_{i} \sum_{i=\text{H,He}}
\int dp'   \frac{\sigma_{i}(p,p')}{dp'} n_{p}(p')
\end{align}
where 
$n_{i}$ is the number density of the interstellar hydrogen (helium),
$n_{p}$ is the number density of primary cosmic-ray proton per total momentum, 
and $d\sigma_{i}(p,p')/dp'$ is the differential cross section
for  $p+\text{H(He)}\to \bar p + X$. 
In calculating the antiprotons, 
inelastic scattering to produce ``tertiary'' antiprotons should be 
taken into account.

The primary source term from 
the annihilation of Majorana DM particles has the  following form
\begin{align}\label{eq:ann-source}
q(\boldsymbol{r},p)=\frac{\rho(\boldsymbol{r})^2}{2 m_{\chi}^2}\langle \sigma v \rangle 
\sum_X \eta_X \frac{dN^{(X)}}{dp} ,
\end{align}
where $\langle \sigma v \rangle$ is 
the velocity-averaged DM annihilation cross section multiplied by DM relative velocity
(referred to as cross section).
$\rho(\boldsymbol{r})$ is the DM energy density distribution function,
and
$dN^{(X)}/dp$ is the injection energy spectrum  of  antiprotons
from DM annihilating into SM final states through 
all possible  intermediate states $X$ with  
$\eta_X$ the corresponding branching fractions.

The   interstellar flux of the cosmic-ray particle is related to its density function as 
$\Phi= v\psi(\boldsymbol{r},p)/(4\pi)$.
At the top of the atmosphere (TOA) of the Earth, 
the fluxes of cosmic-rays  are affected  by solar winds 
and the helioshperic magnetic field. 
This effect is taken into account using 
the force-field approximation
which involves a parameter $\phi$, the  so called Fisk potential
\cite{Gleeson:1968zza}.
In this work,
we shall take  $\phi=550$ MV in the numerical analysis.

We shall solve the diffusion equation of \eq{eq:propagation} 
using the publicly available  code  GALPROP v54
\cite{astro-ph/9807150,astro-ph/0106567,astro-ph/0101068,astro-ph/0210480,astro-ph/0510335}
which utilizes  realistic astronomical information on 
the distribution of interstellar gas and other data as input, 
and considers various kinds of observables
in a self-consistent way. 
Other approaches based on simplified assumptions on 
the Galactic gas distribution 
which  allow  for fast  analytic solutions can be found in Refs.
\cite{astro-ph/0103150,
astro-ph/0212111,
astro-ph/0306207,
1001.0551,
Cirelli:2010xx%
}.

We first consider 
the so-called ``conventional''  diffusive re-acceleration (DR) model 
\cite{astro-ph/0101068,astro-ph/0510335}
which is commonly adopted by the current experimental collaborations 
such as  
PAMELA \cite{Adriani:2008zq,Adriani:2010rc,Adriani:2011xv}  
and 
Fermi-LAT \cite{Ackermann:2010ij,FermiLAT:2012aa}
as a benchmark model for the  astrophysical backgrounds.
It is useful to consider this model as a reference model to 
understand how the DM properties could be  constrained by 
the AMS-02 data.
Then  we consider 
three representative propagation models  selected from 
a large sample of models obtained from 
a global Bayesian MCMC fit to  
the preliminary AMS-02 proton and B/C data using 
the GALPROP code
\cite{Jin:2014ica}.
They are selected  to represent the typically
minimal (MIN), median (MED) and 
maximal (MAX) antiproton fluxes within $95\%$ CL,
corresponding to the region enveloping
$95\%$ of the MCMC samples with highest likelihoods  in 
a six-dimensional parameter space. 
For simple Gaussian posterior distributions, it corresponds to 
$\Delta \chi^{2}=12.59$ for six degrees of freedom.
The parameters in the four models are summarized in 
\tab{tab:prop-models}.

Note that the  ``MIN'', ``MED'' and ``MAX'' models used in this work are
different from and  complementary to 
that given in Ref.
\cite{Donato:2003xg}
in several ways:
\romannumeral 1) 
they  are obtained using the fully numerical GALRPOP code 
which is consistent with the analysis framework of this work 
while  
that in Ref. \cite{Donato:2003xg} are based on 
the two-zone diffusion model with  
the assumption of 
simplified galactic geometry and  uniform interstellar gas distributions.
\romannumeral 2) 
they  correspond to  the DR propagation model
while
that in Ref. \cite{Donato:2003xg} the models with  diffusive-reacceleration plus a constant convection velocity $V_{c}$ are considered.
In the GALPROP approach,
it is known that 
the spectral shape of the B/C flux ratio is better reproduced  without 
including the convection term;
\romannumeral 3) 
the models in this work are based on the global fit to the preliminary AMS-02 data of 
proton flux and B/C flux ratio,
while
the models  in  Ref.~\cite{Donato:2003xg} are based on  
the much older HEAO-3 data on B/C ratio with significantly lower precision.
Consequently, 
the uncertainties  in the prediction for the DM  induced antiproton fluxes
are much larger;
\romannumeral 4)
the Bayesian analysis was used in deriving these models, 
while in  Ref.~\cite{Donato:2003xg} the authors performed a  frequentist analysis.  
The credible intervals obtained from the Bayesian analysis can be different 
from that  using the confidence levels based on a given value of $\Delta \chi^{2}$.

\begin{figure}[tb]
\begin{center}
\includegraphics[width=0.45\textwidth]{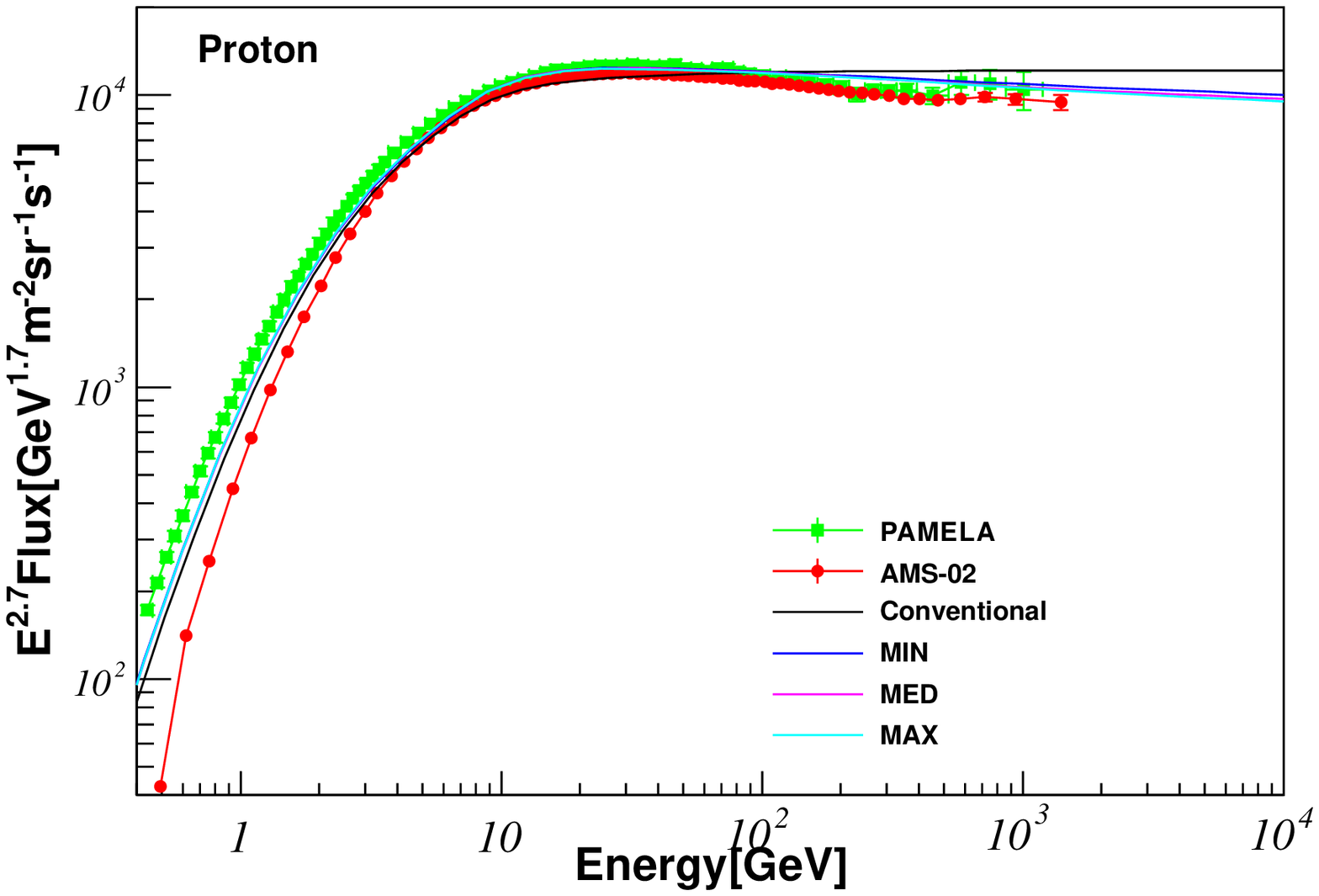}
\includegraphics[width=0.45\textwidth]{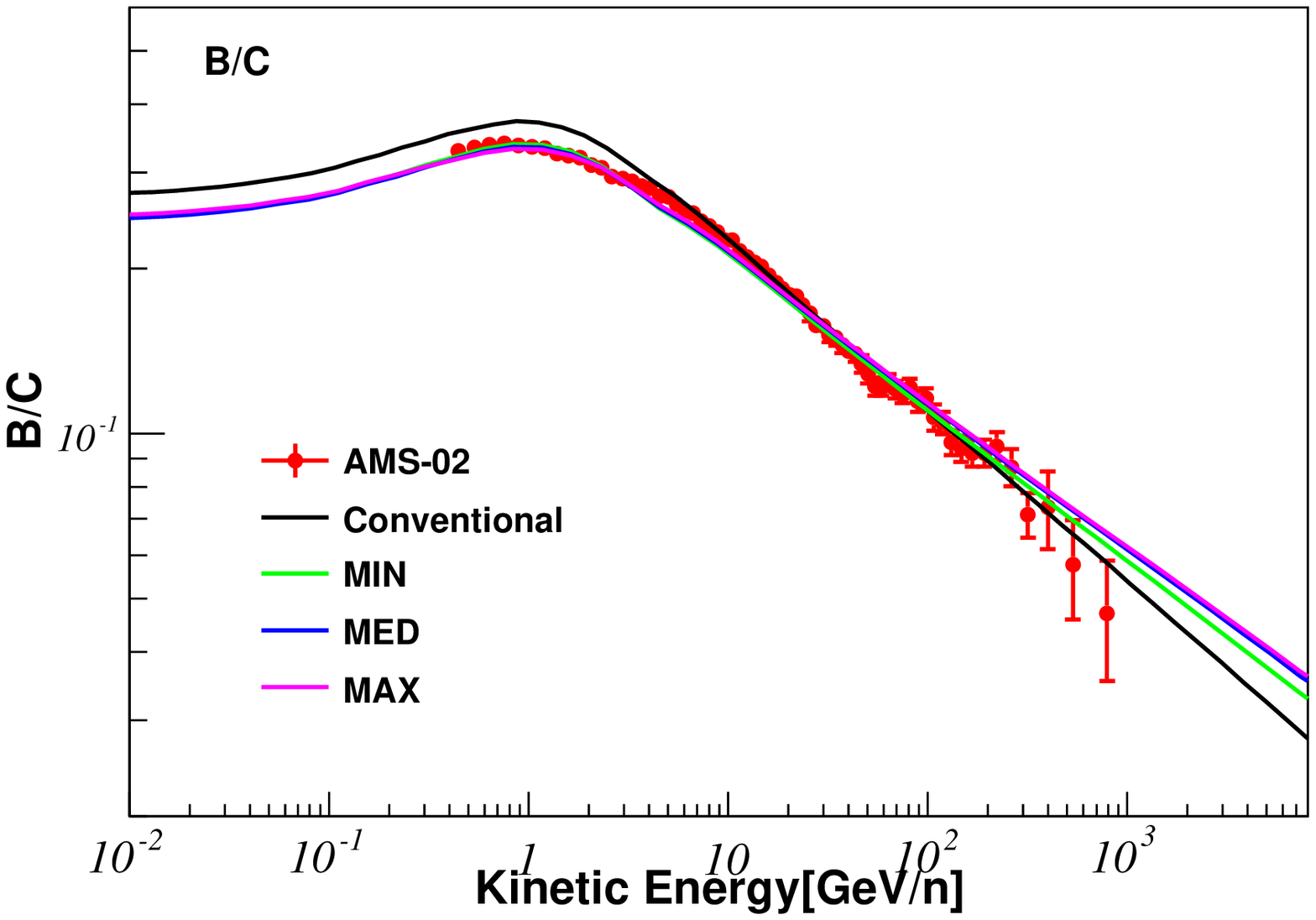}
\caption{
Predictions for the proton flux (left) and the B/C flux ratio (right) in 
the four propagation models listed in \tab{tab:prop-models}.
The latest data of proton flux from AMS-02~\cite{Ting:AMS}
and PAMELA~\cite{
Adriani:2014pza,%
Adriani:2014xoa%
}
are shown.
}
\label{fig:protonBC}
\end{center}
\end{figure}

\begin{figure}[tb]
\begin{center}
\includegraphics[width=0.65\textwidth]{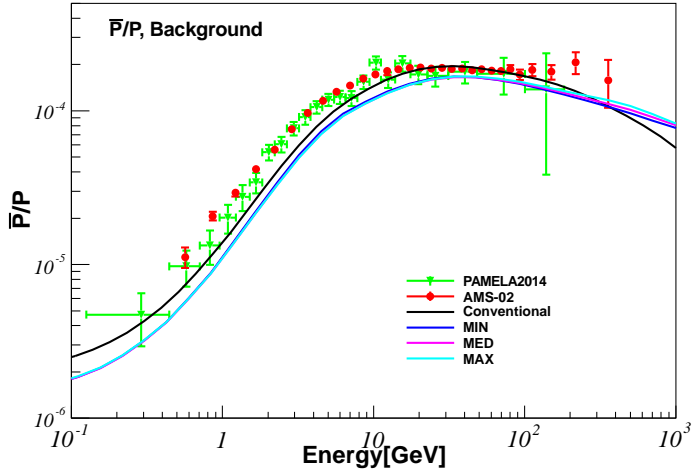}
\caption{
Predictions for the $\bar p /p$ ratio from the four propagation models
list in \tab{tab:prop-models}. 
The data  from AMS-02~\cite{Ting:AMS}
and PAMELA~\cite{Adriani:2014pza} are shown.
}
\label{fig:pbar}
\end{center}
\end{figure}

The predicted proton flux and the B/C flux ratio in these models are shown in \fig{fig:protonBC}, 
together with the latest AMS-02 data
\cite{Ting:AMS}.
The figures show an overall agreement with the current data in these models. 
In the ``conventional'' model, 
the predicted B/C ratio is a little higher for the kinetic energy below $\sim10$ GeV/n,
but are consistent with the B/C data in the higher energies.
Note that  the theoretical predictions  from  Ref.~\cite{Jin:2014ica}
is based on  the analysis of 
the previous preliminary AMS-02 results announced at the 
conference ICRC(2013)
\cite{
Haino:icrc2013,
Oliva:icrc2013%
}.
So far the AMS-02 data on the B/C flux ratio is still unpublished. 
We  will update the whole analysis in Ref.~[26] after they get published, 
in a future work. 
The predictions for the background of the $\bar p/p$ flux ratio in these models are 
shown in \fig{fig:pbar}.
The ``MIN'', ``MED'' and ``MAX'' models are highly degenerate in the background 
$\bar p/p$ ratio.
Compared with these models, 
the ``conventional'' model predicts more low energy antiprotons but 
at high energies above $\sim 500$~GeV, the predicted antiprotons are much less. 
In all the four DR propagation models, below $\sim10$~GeV 
the GALPROP based calculations  underpredict
the  $\bar p/p$ flux ratio by $\sim 40\%$, which is a known issue.
The agreement with the low energy $\bar p$ data can be improved by
introducing breaks in diffusion coefficients
\cite{Moskalenko:2001ya}, 
``fresh'' nuclei component
\cite{Moskalenko:2002yx} %
or a DM contribution
\cite{Hooper:2014ysa}. %
Note that 
at energies below $\sim 10$~GeV, 
the uncertainties due to  
the  solar modulation and the propagation parameters  increase significantly.
The predictions for low energy $\bar p/p$ ratio can be easily modified  by 
introducing an independent Fisk potential $\phi$ for $\bar p$ and  
an energy-dependent overall normalization factor as 
discussed  in Ref.\cite{Giesen:2015ufa}.
In this  work, 
we instead use these DR models to 
derived  very conservative upper limits on 
the annihilation cross sections of light DM particles.
Note however that in the DR propagation models,
the background predictions agree with the AMS-02 data well
at higher energies $\sim 10-100$ GeV,
which  can be turned into stringent constraints on 
the nature of heavy DM particles.
\begin{table}[htb]
  \begin{center}
  \begin{tabular}{lllllllll}\hline\hline
    model &$R (\mbox{kpc})$& $Z_{h}(\mbox{kpc})$ & $D_{0}$ & $\rho_{0}$&$\delta_{1}/\delta_{2}$   & $V_{a}(\mbox{km}/\mbox{s})$ &$\rho_{s}$&$\gamma_{p1}/\gamma_{p2}$ \\
 \hline
Conventional           
&20   & 4.0   & 5.75 &4.0  & 0.34/0.34     & 36.0  & 9.0  & 1.82/2.36 \\
MIN       
&20   & 1.8   & 3.53  &4.0  &  0.3/0.3        &42.7 &10.0 & 1.75/2.44\\
MED        
&20   & 3.2   & 6.50  &4.0  &  0.29/0.29    &44.8  &10.0 & 1.79/2.45\\
MAX        
&20   & 6.0   & 10.6 &4.0  &  0.29/0.29      &43.4 &10.0 & 1.81/2.46\\
  \hline\hline
 \end{tabular}\end{center}
\caption{
Parameters in the propagation models
``Conventional''
\cite{astro-ph/0101068,astro-ph/0510335}, 
``MIN'', ``MED'' and ``MAX'' models from 
Ref.~\cite{Jin:2014ica}.
 $D_{0}$ is in units of  $10^{28}\mbox{cm}^{2}\cdot\mbox{s}^{-1}$, 
 the break rigidities $\rho_{0}$ and $\rho_{s}$  are in units of GV.}
 \label{tab:prop-models}
\end{table}

The flux cosmic-ray antiprotons from DM annihilation depend also significantly on 
the choice of DM halo profile.  
N-body simulations suggest a universal form of the DM profile 
\begin{align}
\rho(r)=\rho_\odot
\left( \frac{r}{r_\odot} \right)^{-\gamma}
\left(
\frac{1+(r_\odot/r_s)^\alpha}{1+(r/r_\odot)^\alpha}
\right)^{(\beta-\gamma)/\alpha} ,
\end{align}
where 
$\rho_\odot \approx 0.43 \text{ GeV}\text{ cm}^{-3}$ is
the local DM energy density 
\cite{Salucci:2010qr%
}.
The values of the parameters $\alpha$, $\beta$, $\gamma$ and $r_{s}$
for the  Navarfro-Frenk-White (NFW) profile~\cite{Navarro:1996gj},
the isothermal profile~\cite{Bergstrom:1997fj} 
and the Moore profile 
\cite{Moore:1999nt, %
Diemand:2004wh} 
are summarized in \tab{tab:DMprofile}.
\begin{table}[htb]
\begin{center}
\begin{tabular}{ccccc}
\hline\hline
             	&$\alpha$& $\beta$& $\gamma$& $r_{s}$(kpc)\\
\hline
NFW		& 1.0 	& 3.0	& 1.0 	& 20\\ 
Isothermal& 2.0	&2.0		& 0 		& 3.5\\
Moore	 & 1.5	&3.0		&1.5		&28.0\\
\hline\hline
\end{tabular}
\caption{
Values of parameters $\alpha$, $\beta$, $\gamma$ and $r_{s}$ for three
DM halo models,
NFW 
\cite{Navarro:1996gj}, %
Isothermal 
\cite{Bergstrom:1997fj}, %
and 
Moore 
\cite{Moore:1999nt, %
Diemand:2004wh}. %
}
\label{tab:DMprofile}
\end{center}
\end{table}
An other widely adopted  DM profile is the  Einasto profile 
\cite{Einasto:2009zd%
}
\begin{align}
\rho(r)=\rho_\odot \exp
\left[
-\left( \frac{2}{\alpha_E}\right)
\left(\frac{r^{\alpha_E}-r_\odot^{\alpha_E}}{r_s^{\alpha_E}} \right)
\right] ,
\end{align}
with $\alpha_E\approx 0.17$ and $r_s\approx 20$ kpc. 

We consider three reference DM annihilation channels 
$\bar\chi\chi \to XX$ 
where $XX=q \bar q$, $b \bar b$ and $W^{+}W^{-}$.
The energy spectra of these channels are similar at high energies.
The main difference is in the average number of total antiprotons $N_{X}$
per DM annihilation of each channel.
For a  DM particle mass $m_{\chi}=500$ GeV, 
the values of $N_{X}$ for typical final states are 
$N_{q\bar q} = 2.97~(q=u,d)$,
$N_{b\bar b}= 2.66$,
and
$N_{WW}=1.42$.
The injection spectra $dN^{(X)}/dp$ from DM annihilation are calculated using 
the numerical package PYTHIA~v8.175
\cite{%
Sjostrand:2007gs%
}.
in which the long-lived particles such as neutron and $K_{L}$ are allowed to decay
and the final state interaction are taken into account.
Since PYTHIA~v8.15 the polarization and correlation of final states in  $\tau$-decays 
has been  taken into account
\cite{Ilten:2012zb}.

\begin{figure}[tb]
\begin{center}
\includegraphics[width=0.45\textwidth]{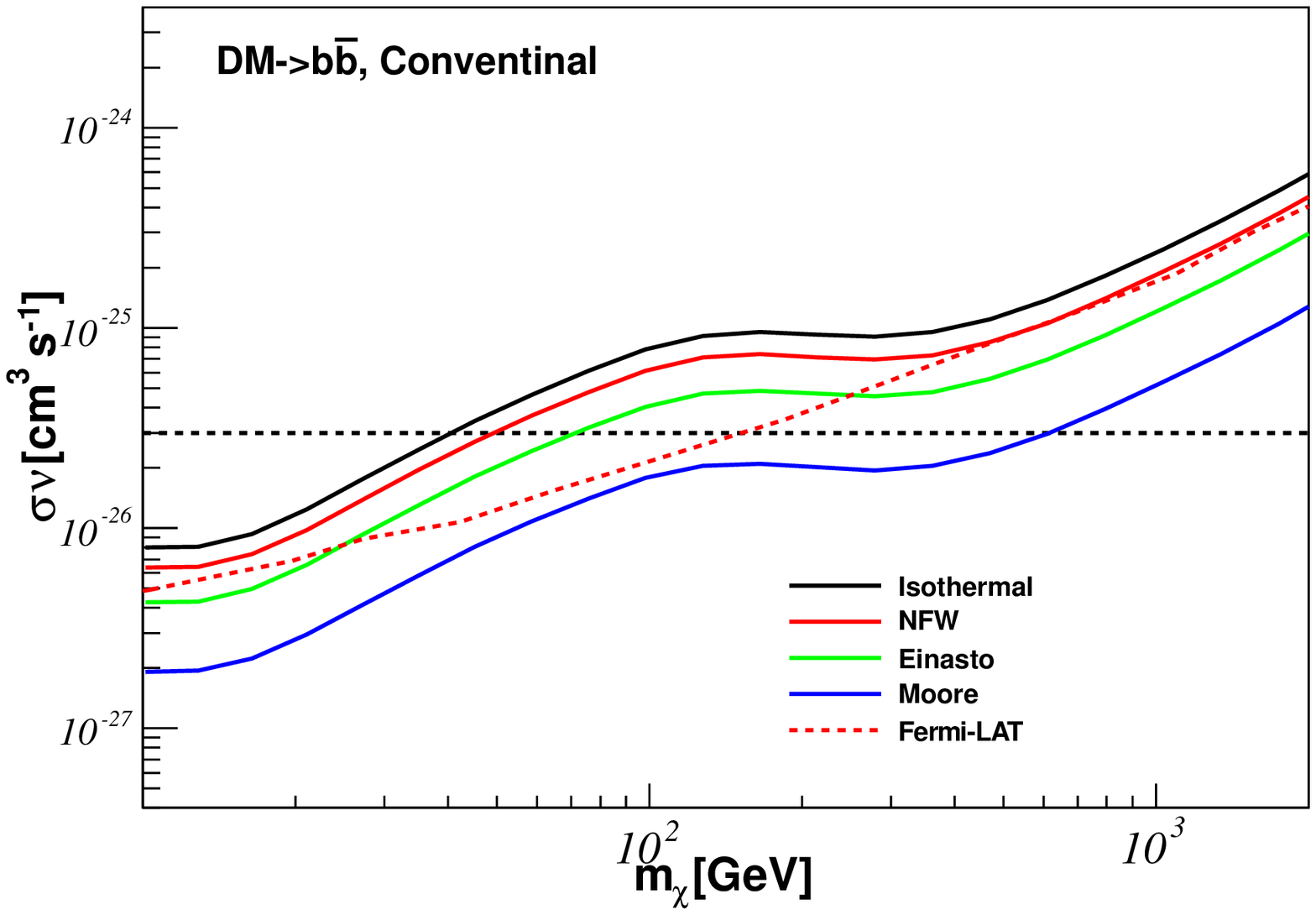}
\includegraphics[width=0.45\textwidth]{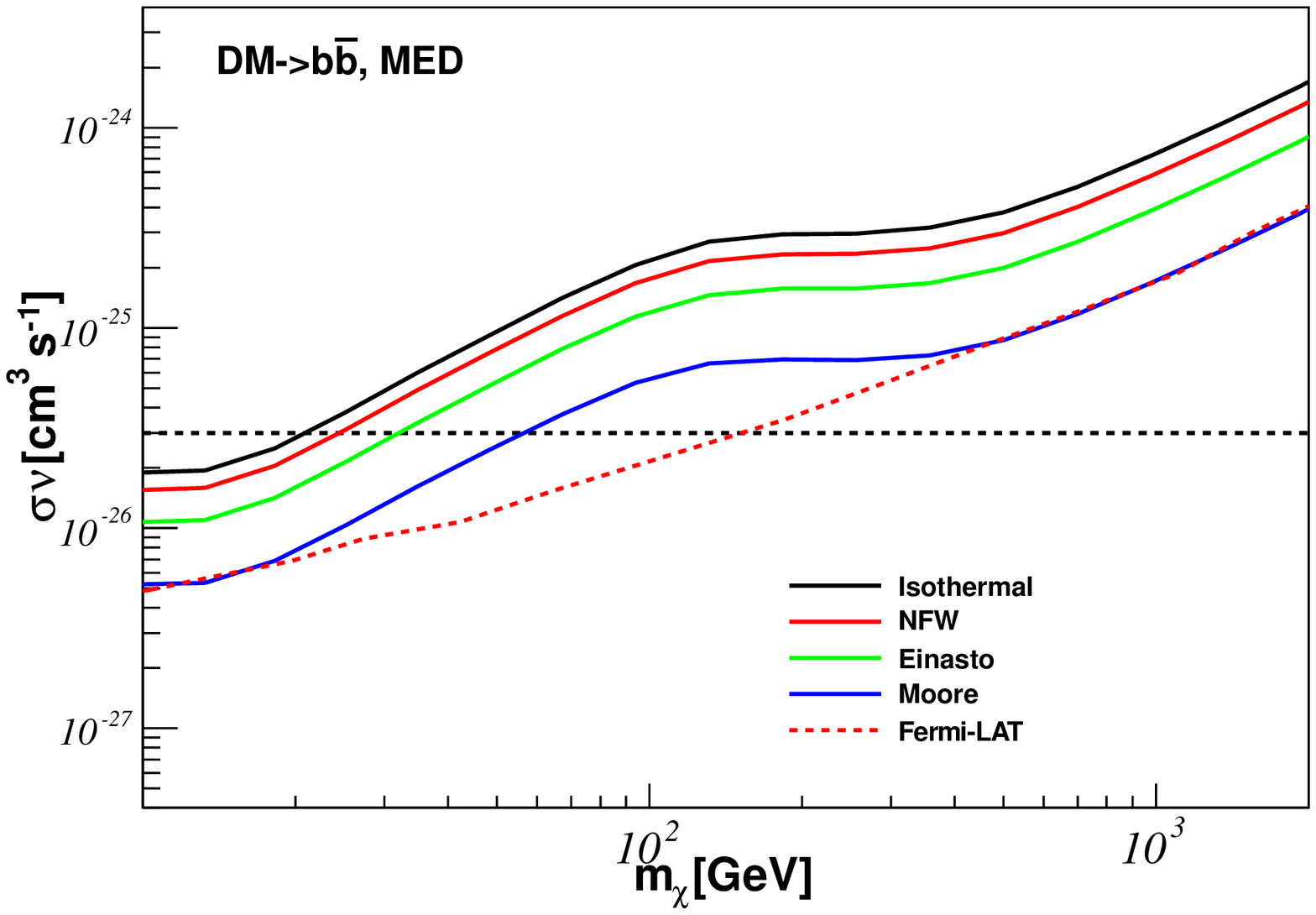}
\\
\includegraphics[width=0.45\textwidth]{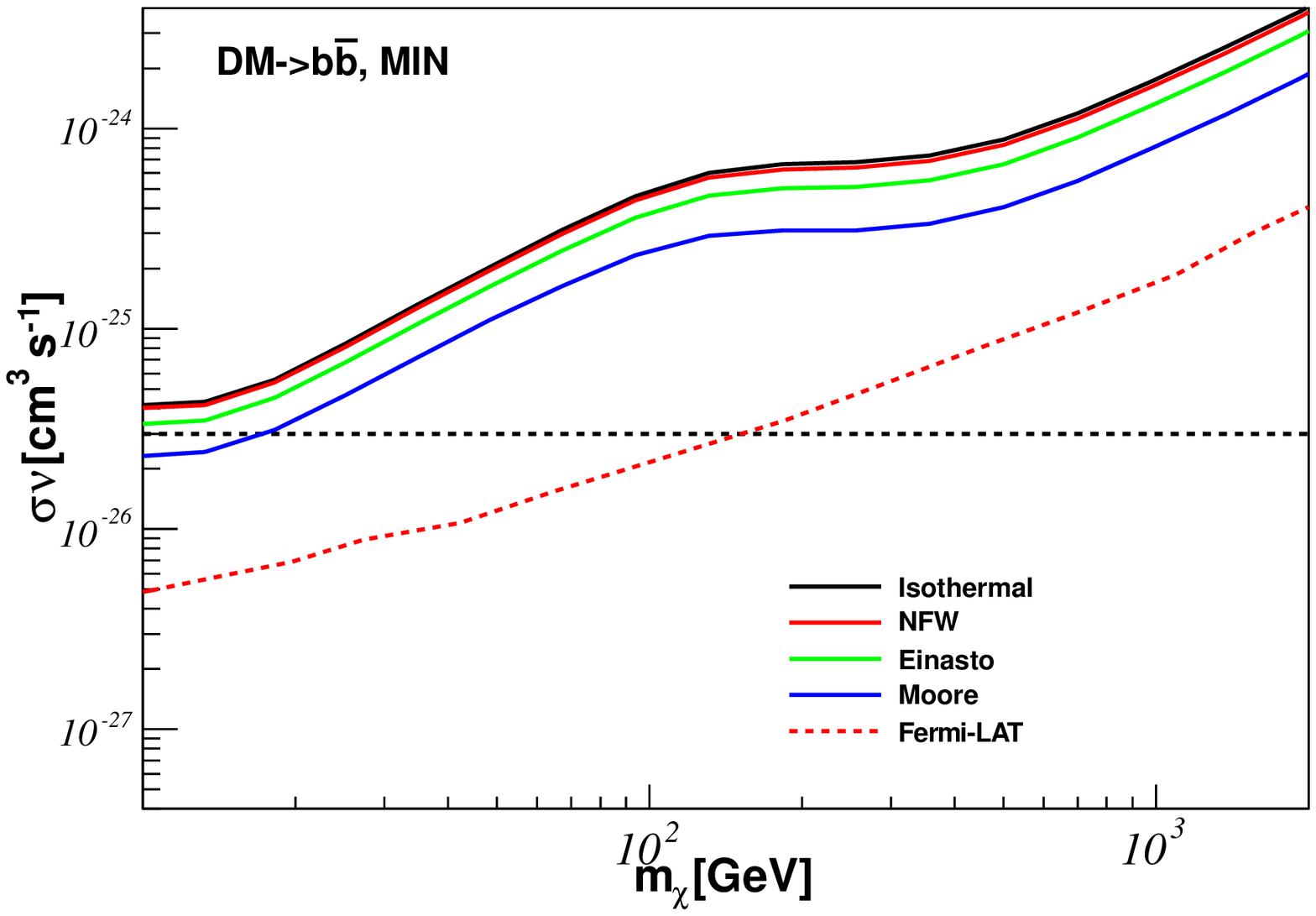}
\includegraphics[width=0.45\textwidth]{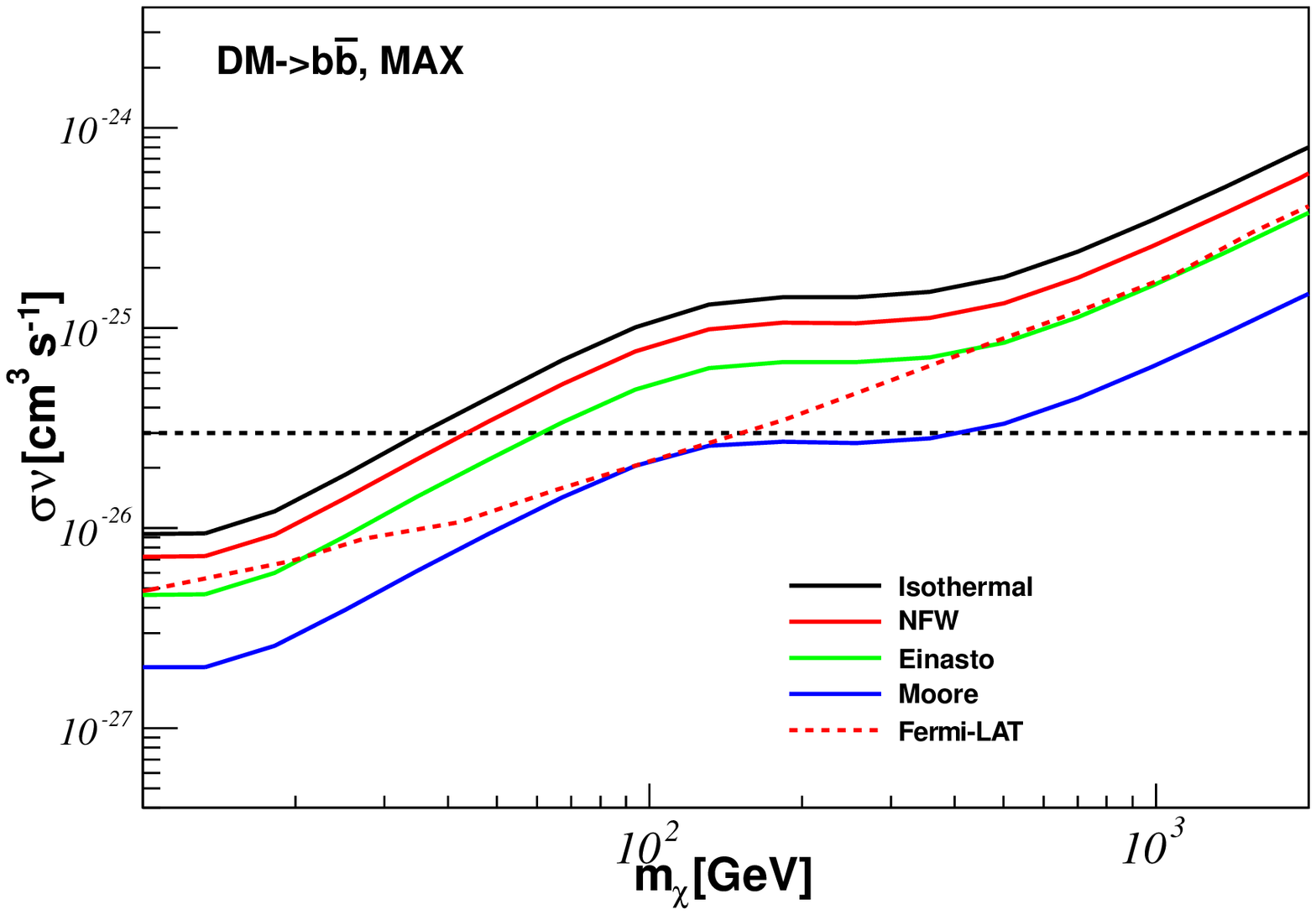}
\caption{
Upper limits on 
the cross sections for 
DM particle annihilation into $b\bar b$ final states from
the AMS-02 $\bar p/p$ data in 
the ``conventional'' (upper left),
``MED'' (upper right),
``MIN'' (lower left) and
``MAX'' (lower right)  propagation models.
Four DM profiles 
NFW~\cite{Navarro:1996gj}, 
Isothermal~\cite{Bergstrom:1997fj} , 
Einasto 
\cite{Einasto:2009zd%
}
and Moore 
\cite{Moore:1999nt, %
Diemand:2004wh} 
are considered.
The upper limits  from 
the Fermi-LAT 6-year gamma-ray data of 
the dwarf spheroidal satellite galaxies of the Milky Way
are also shown
\cite{Ackermann:2015zua}.
The horizontal line indicates the typical thermal annihilation cross section
$\langle \sigma v \rangle=3\times 10^{-26}\text{cm}^{3}\text{s}^{-1}$.
}
\label{fig:bb}
\end{center}
\end{figure}

 In this work,
we shall first derive the upper limits on the DM annihilation cross section 
as a function of DM particle mass,
using the frequentist $\chi^{2}$-analyses.
 The expression of $\chi^2$ is defined as
\begin{align}
\chi^2=\sum_i \frac{(f_i^{\text{th}}-f_i^{\text{exp}})^2}{\sigma_i^2},
\end{align} 
where $f_i^{\text{th}}$ are the theoretical predictions. 
$f_i^{\text{exp}}$ and $\sigma_i$ are the central values  and errors of experimental data, respectively. 
The index $i$ runs over all the available data points.
For a given DM particle mass, 
we first calculate the minimal value $\chi_{\text{min}}^{2}$ of the $\chi^2$-function,
and 
then derive the one-side  $95\%$ CL upper limits on the annihilation cross section,
corresponding to  $\Delta \chi^{2}=3.84$ for one parameter.
All of the 30 data points of the AMS-02 $\bar p/p$ data are included in calculating 
the limits.

\begin{figure}[tb]
\begin{center}
\includegraphics[width=0.45\textwidth]{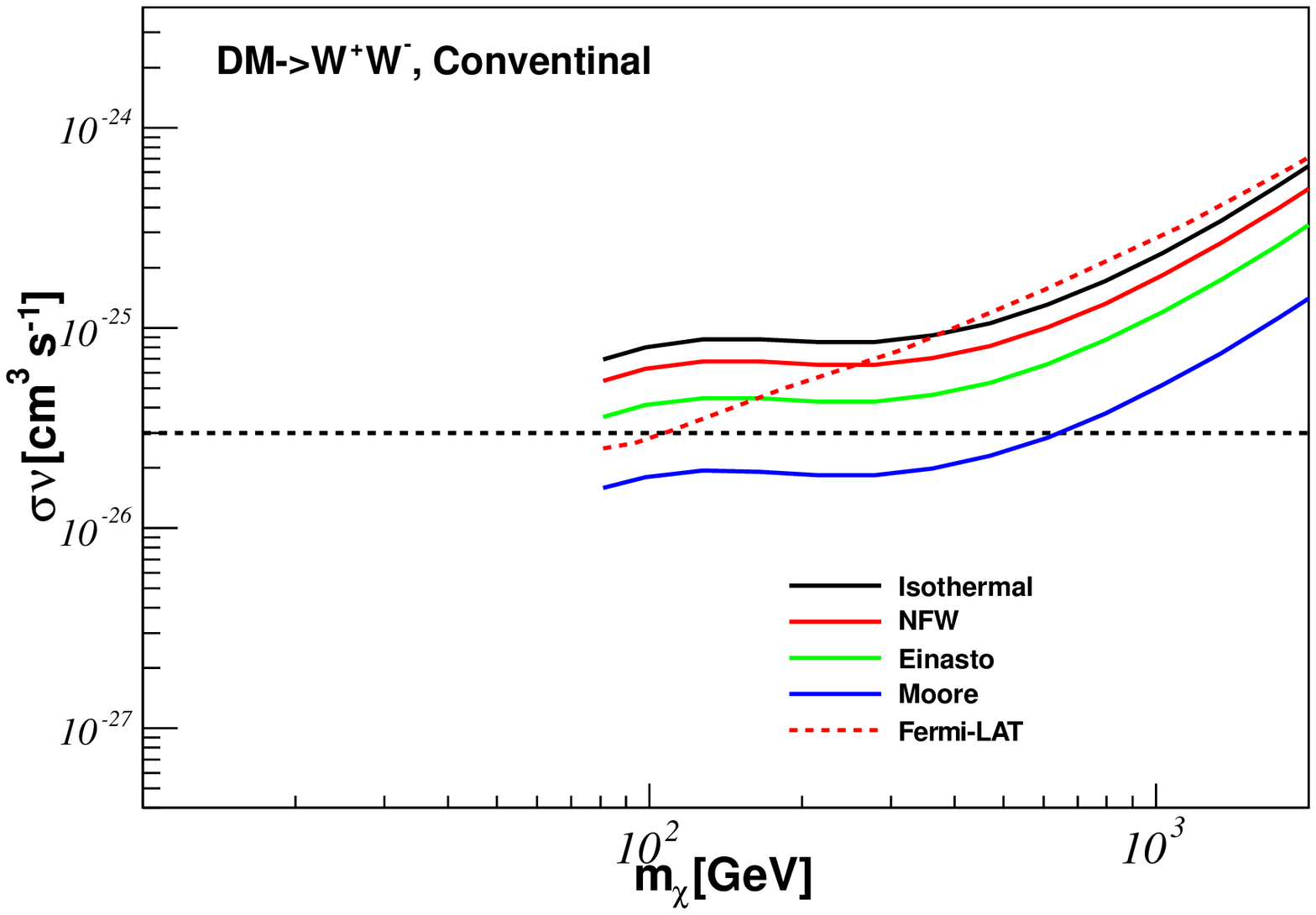}
\includegraphics[width=0.45\textwidth]{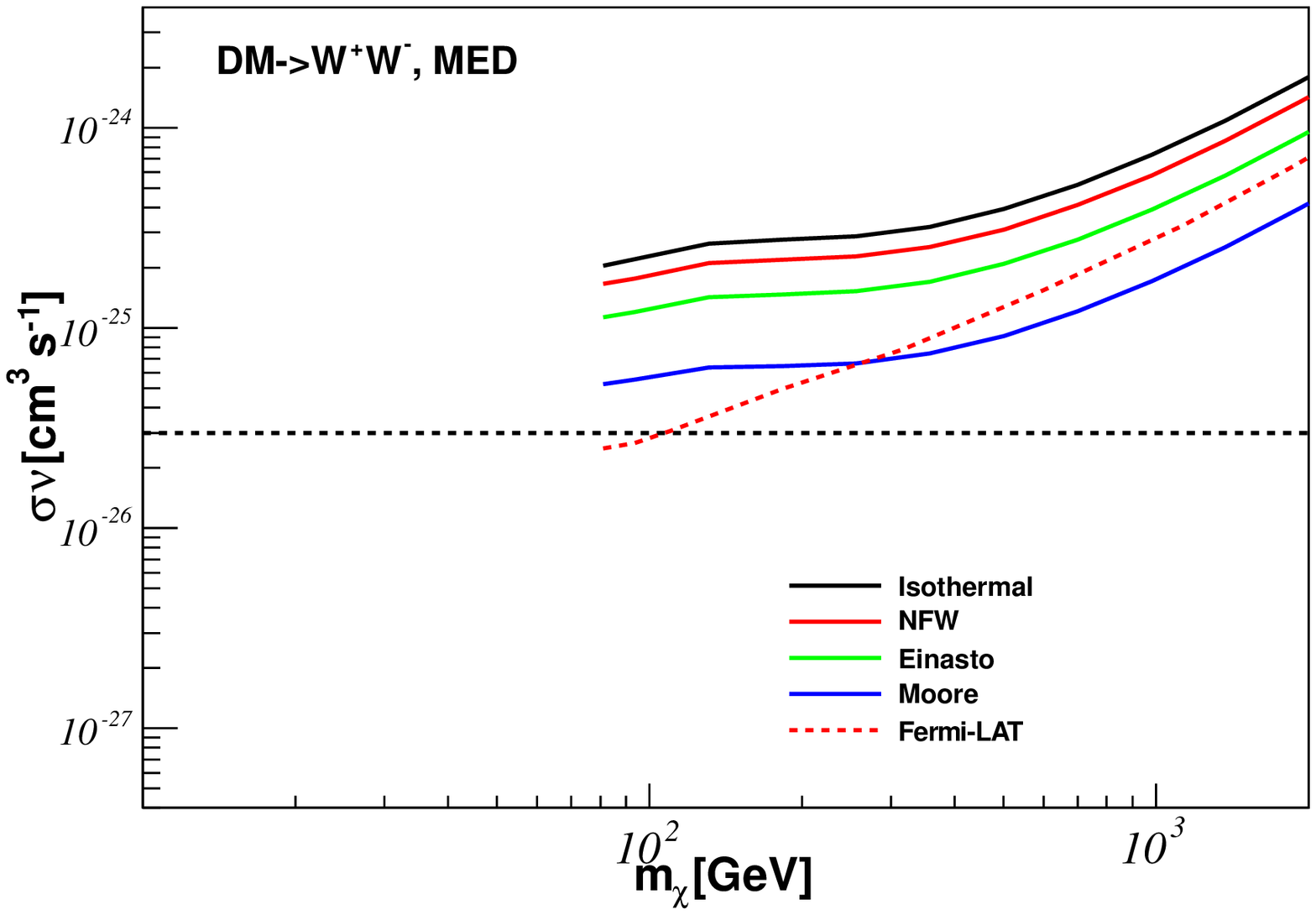}
\\
\includegraphics[width=0.45\textwidth]{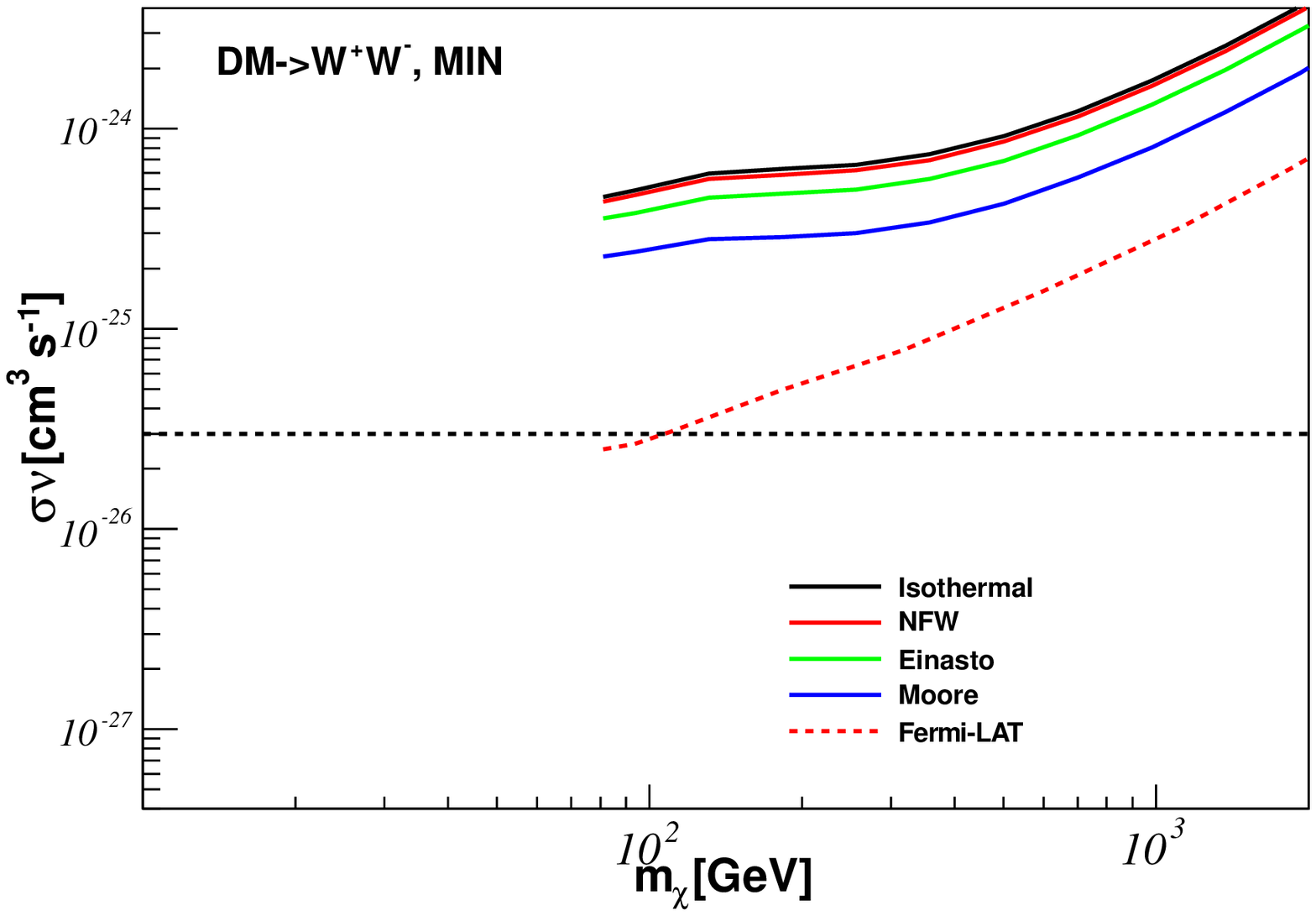}
\includegraphics[width=0.45\textwidth]{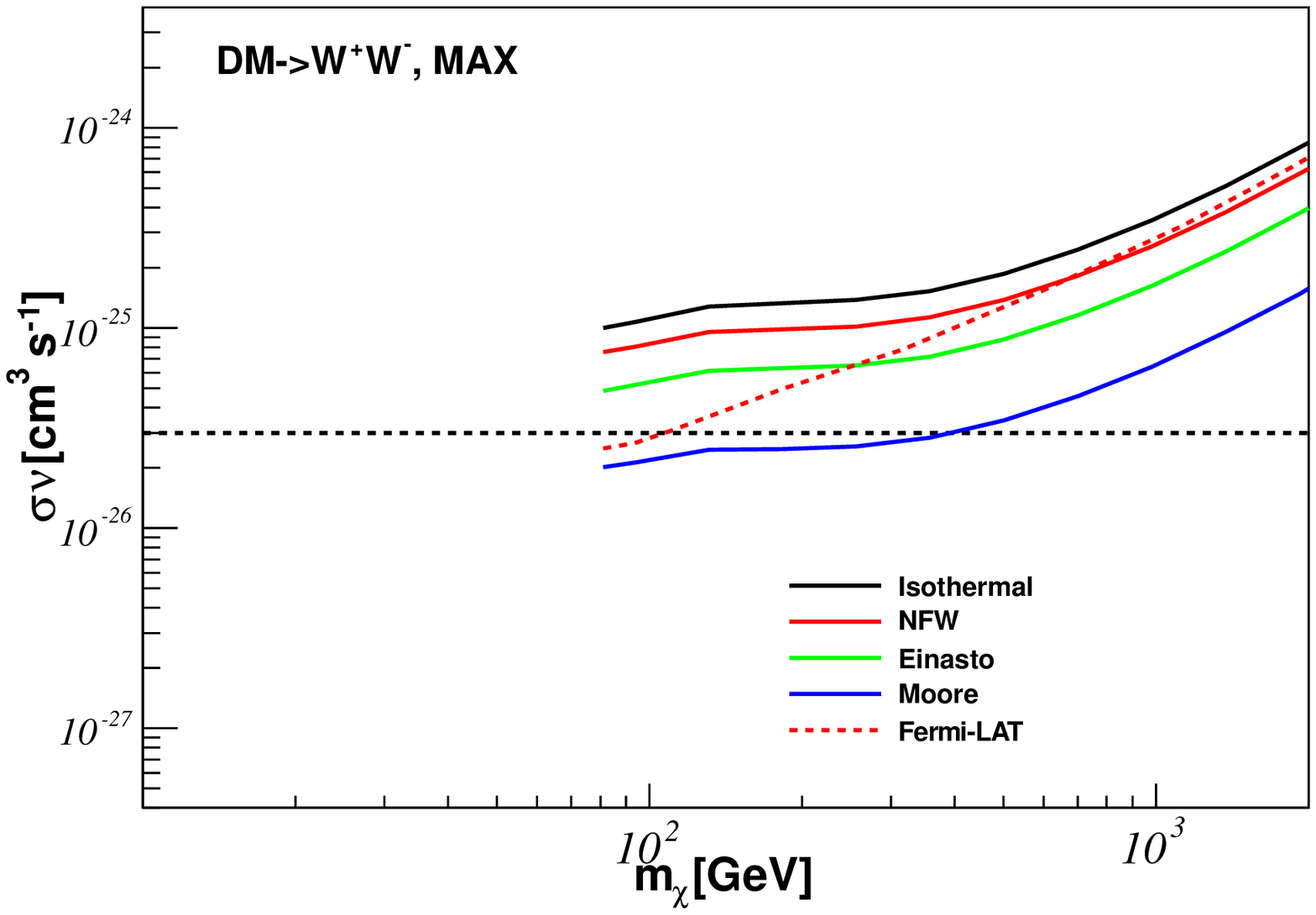}
\caption{The same as \fig{fig:bb}, but for DM annihilation into $W^{+}W^{-}$ final states.}
\label{fig:WW}
\end{center}
\end{figure}

In \fig{fig:bb}, 
we show the obtained upper limits on 
the cross sections for 
DM particle annihilation into $b\bar b$ final states from
the AMS-02 $\bar p/p$ data in 
the ``conventional'', ``MED'',  ``MIN''  and ``MAX'' propagation models.
Four different DM profiles 
NFW~\cite{Navarro:1996gj}, 
Isothermal~\cite{Bergstrom:1997fj}, 
Einasto~\cite{Einasto:2009zd%
}
and 
Moore~\cite{
Moore:1999nt, %
Diemand:2004wh} 
are considered.
As can be seen, 
the upper limits as a function of $m_{\chi}$ show some smooth structure for
all the final states and DM profiles.
The limits tend to be relatively stronger at $m_{\chi}\approx 300$ GeV,
which is related to the fact that the background predictions agree with 
the data well at the antiproton energy range $\sim 20-100$ GeV. 
For a comparison,
the upper limits  from 
the Fermi-LAT 6-year gamma-ray data of 
the dwarf spheroidal satellite galaxies of the Milky Way
are also shown
\cite{Ackermann:2015zua}.
In the ``conventional'' model, 
the upper limits from the AMS-02 $\bar p/p$ data are found to be compatible with that derived from
the Fermi-LAT gamma-ray data for $m_{\chi}\gtrsim 300$ GeV. 
This observation holds for most of the DM profiles.
In the ``MED'' model, the constraints are relatively weaker,
which is related to the under prediction of low energy antiprotons
in this model and the limits are more conservative.
For an estimation of the uncertainties due to the propagation models,
from the ``MIN'' model to the ``MAX'' model, 
we find that the variation of the upper limits is within about a factor of five.

\begin{figure}[tb]
\begin{center}
\includegraphics[width=0.45\textwidth]{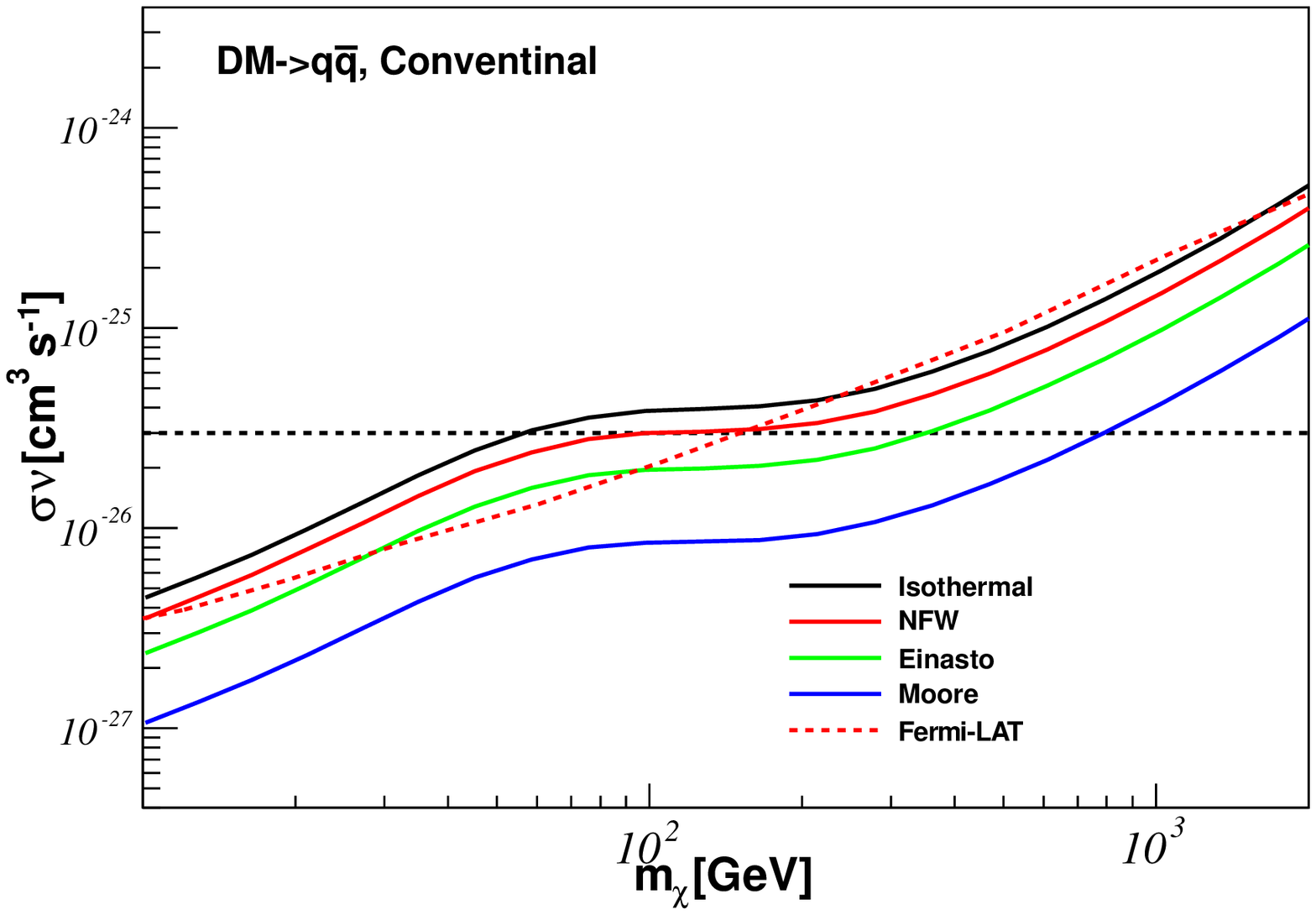}
\includegraphics[width=0.45\textwidth]{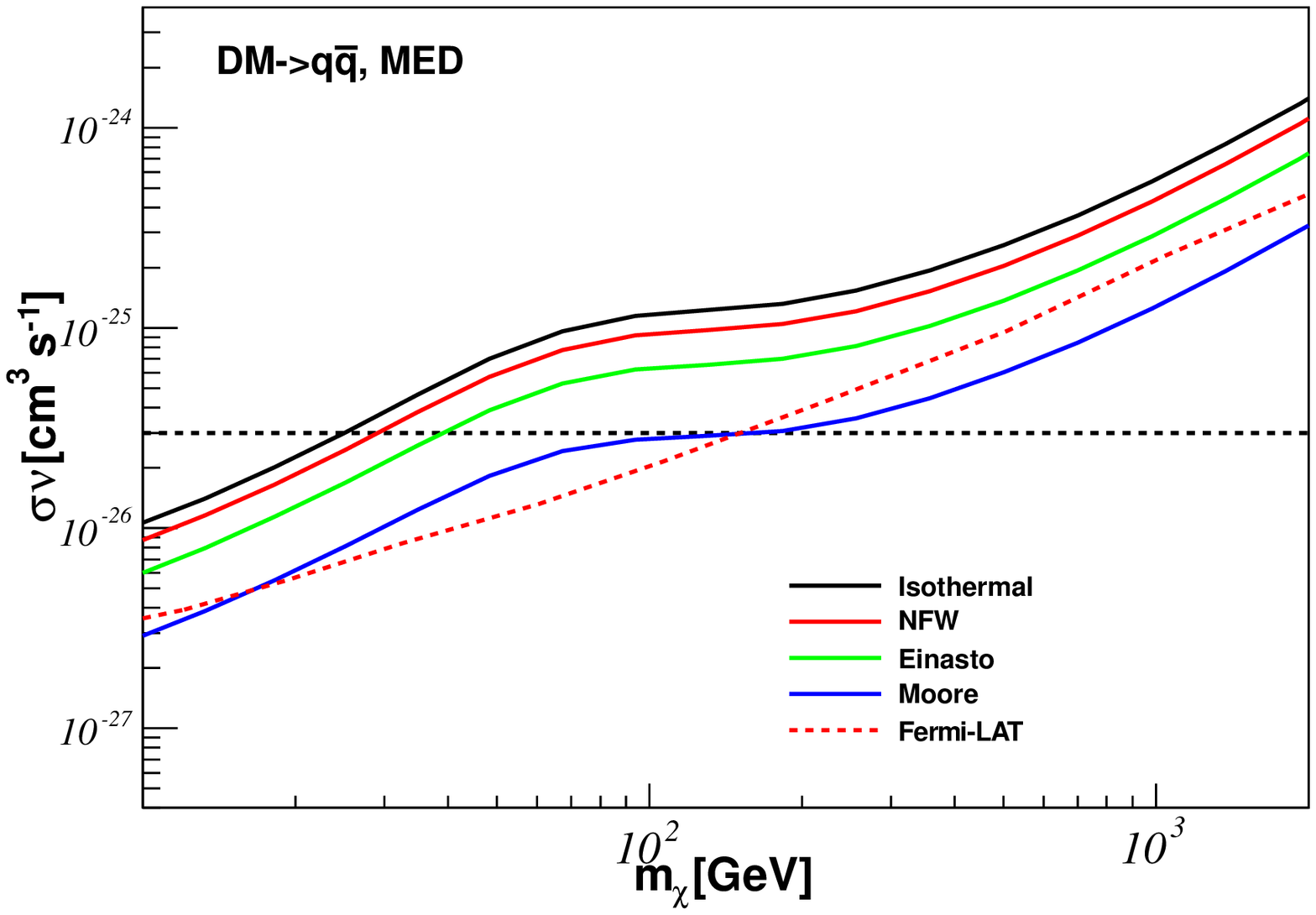}
\\
\includegraphics[width=0.45\textwidth]{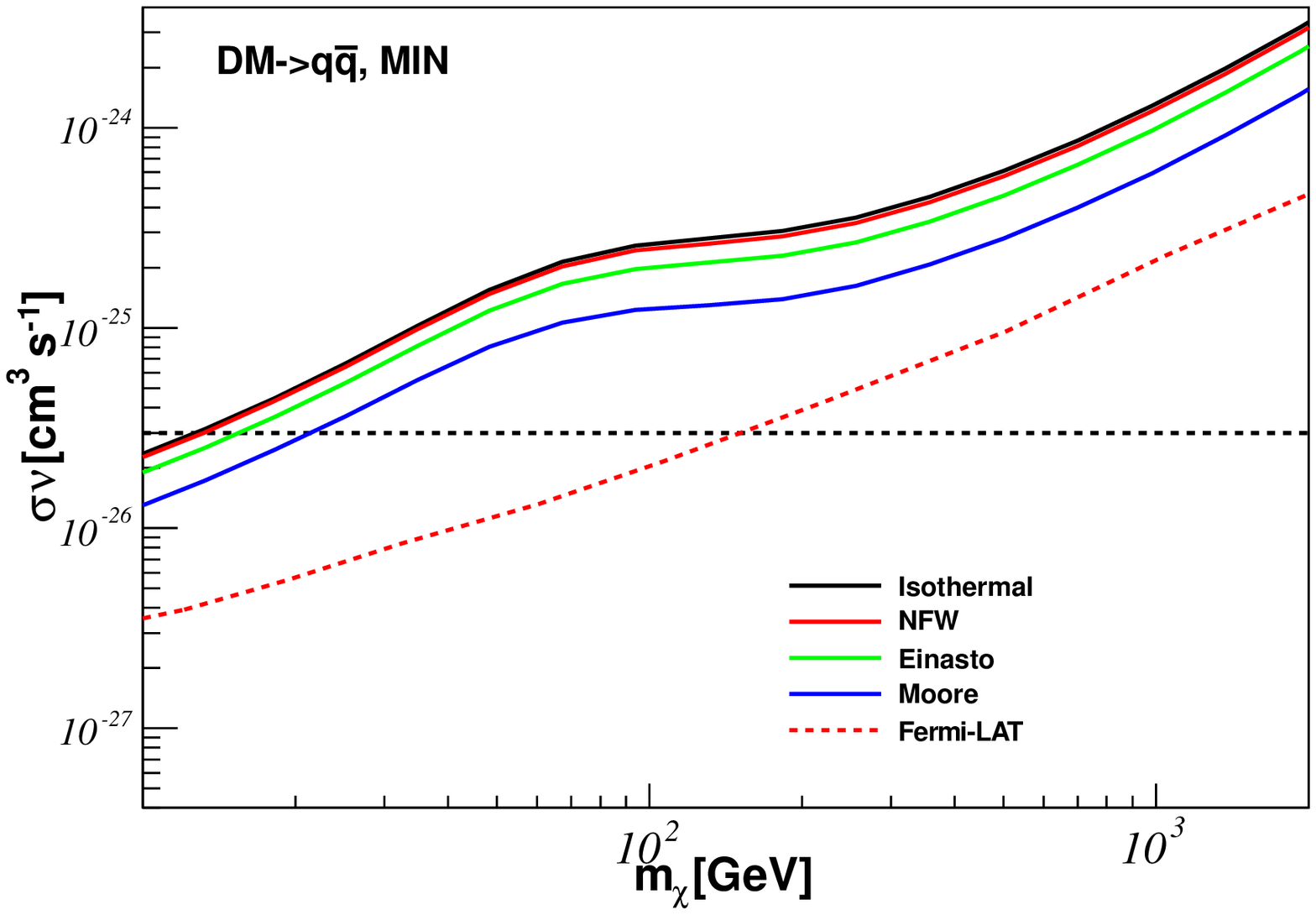}
\includegraphics[width=0.45\textwidth]{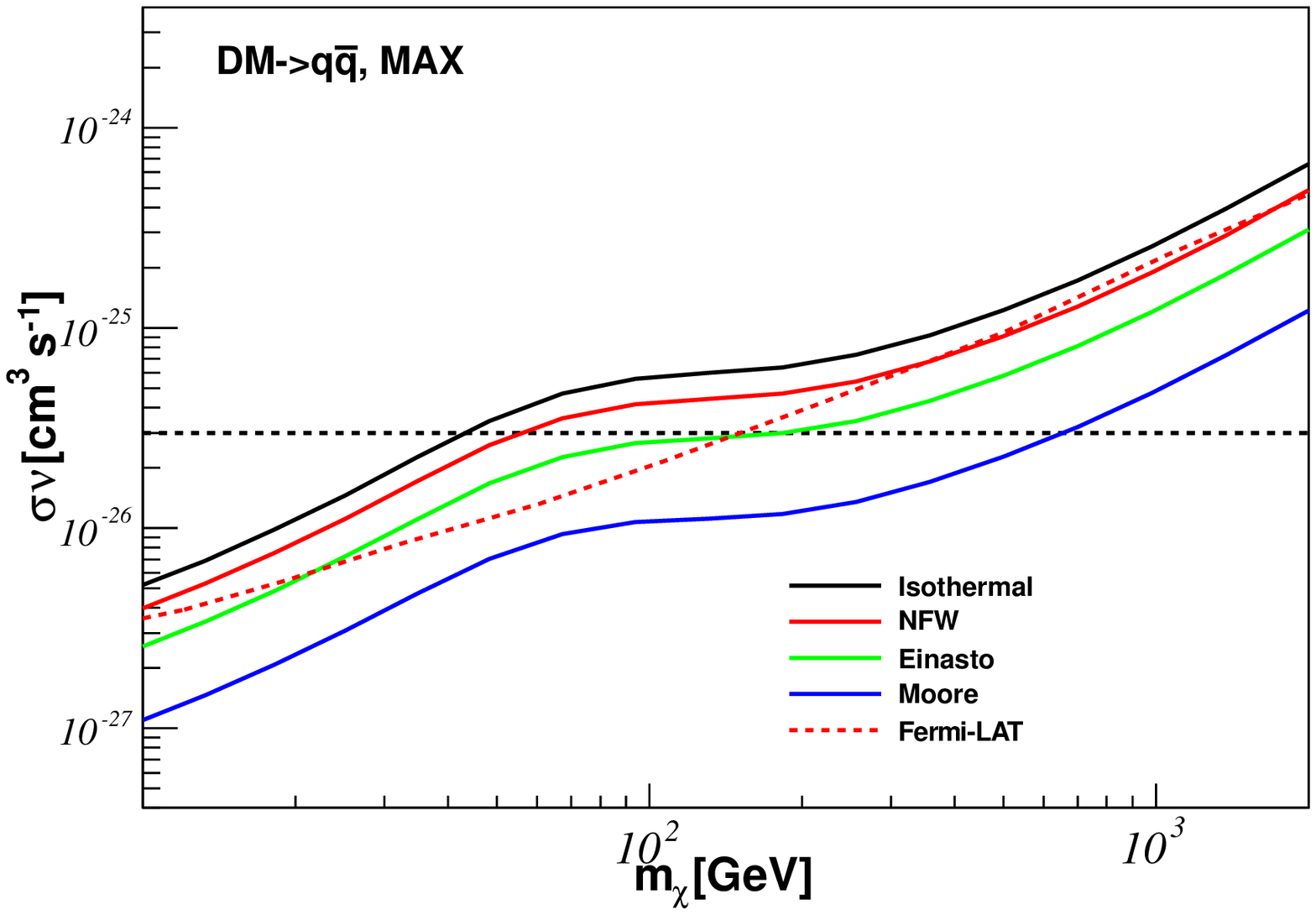}
\caption{The same as \fig{fig:bb}, but for DM annihilation into $q\bar q$ final states.}
\label{fig:qq}
\end{center}
\end{figure}

For the $W^{+}W^{-}$ final states, 
the results are shown in \fig{fig:WW}.
The constraints from AMS-02 $\bar p/p$ data turn out to be
more stringent than that from 
the Fermi-LAT gamma-ray data for 
all the four DM profiles in
the ``convention'' model when the DM particle mass is above $\sim300$ GeV.
Again we find that the variation of the upper limits from the ``MIN'' to the ``MAX''
model is within a factor of five.
The result for the $q\bar q$ final states is shown in \fig{fig:qq}.
Similar to the case of $W^{+}W^{-}$ final states, 
the constraints from $\bar p/p$ data are more stringent at 
about $\sim300$ GeV.
Compared with the case of $W^{+}W^{-}$ and $b\bar b$, 
the constraints on the  $q\bar q$ final states are the most stringent.
For all the three final states, 
we find that 
the allowed DM annihilation cross section is below 
the typical thermal cross section for $m_{\chi}\lesssim 300$ GeV in 
the conventional propagation model with Einasto profile,
which shows that the AMS-02 $\bar p/p$ data can impose 
stringent constraints on DM candidates of weakly interacting massive particles.

\begin{figure}[tbh]
\begin{center}
\includegraphics[width=0.45\textwidth]{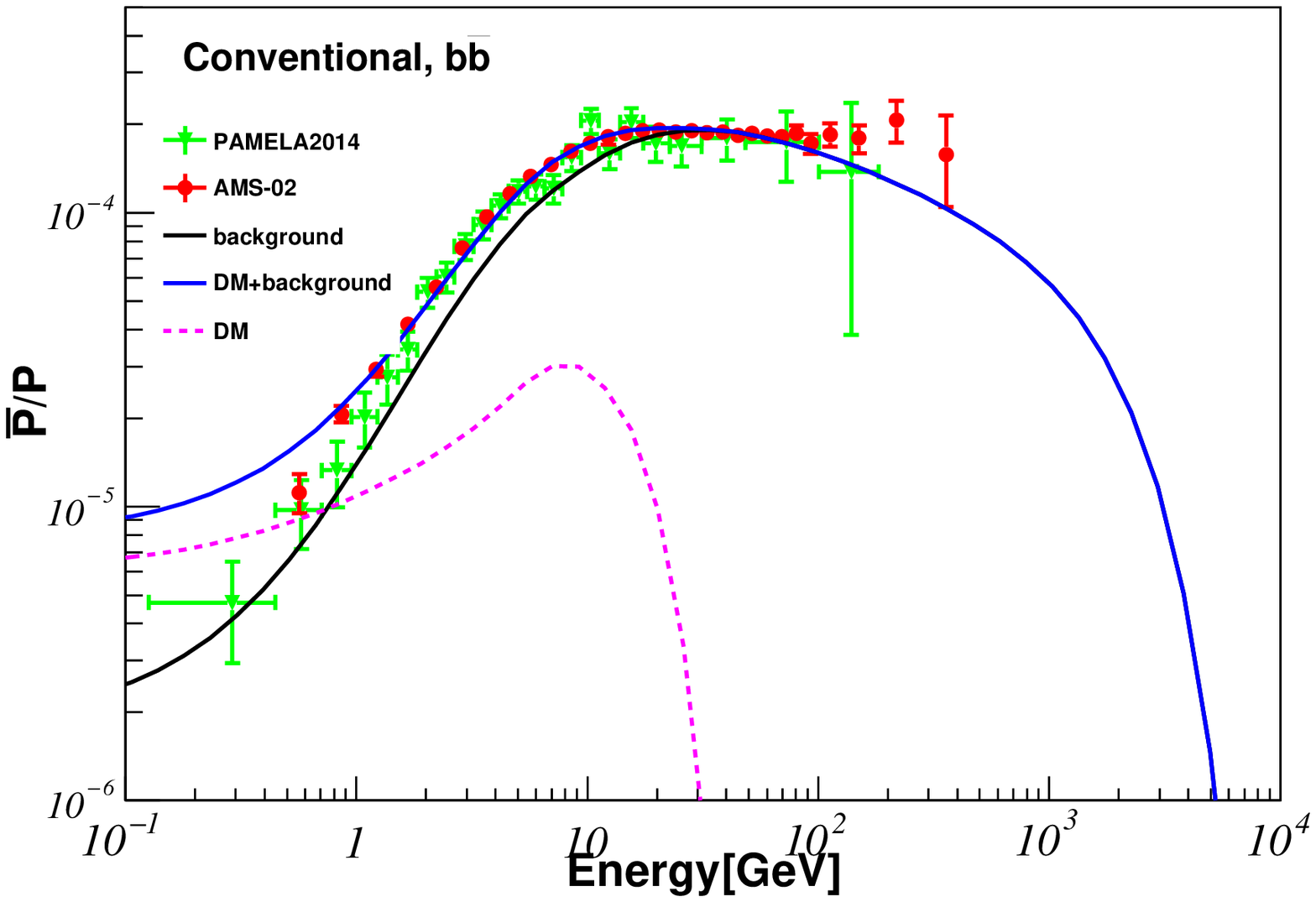}
\includegraphics[width=0.45\textwidth]{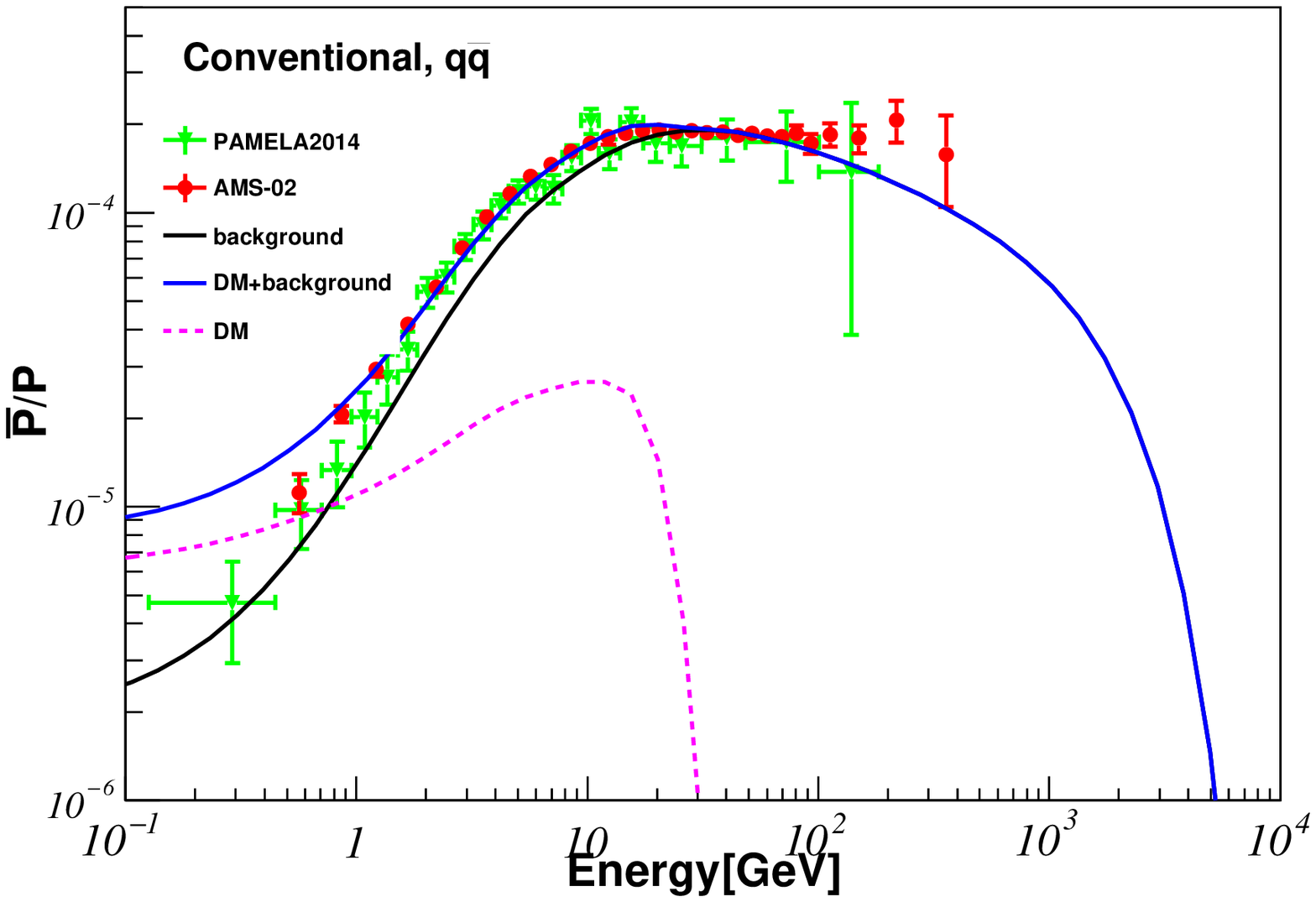}
\caption{ 
Left) 
Spectrum of  $\bar p/p$ flux ratio  from DM annihilating into  $\bar b b$ final states with
$m_{\chi}=58.5 $~GeV and 
$\langle \sigma v \rangle=2.16\times 10^{-26}\text{ cm}^{3}\text{s}^{-1}$ obtained from
a fit to the whole AMS-02  $\bar p/p$ data~\cite{Ting:AMS}. 
The ``conventional'' background model and the Einasto DM profile are assumed.
Right) The same as left, but for the fit with $\bar q q$ final state with 
the best-fit values
$m_{\chi}=35$~GeV and 
$\langle \sigma v \rangle=0.86\times 10^{-26}\text{ cm}^{3}\text{s}^{-1}$.
}
\label{fig:low-energy-fit}
\end{center}
\end{figure}

\begin{figure}[htb]
\begin{center}
\includegraphics[width=0.45\textwidth]{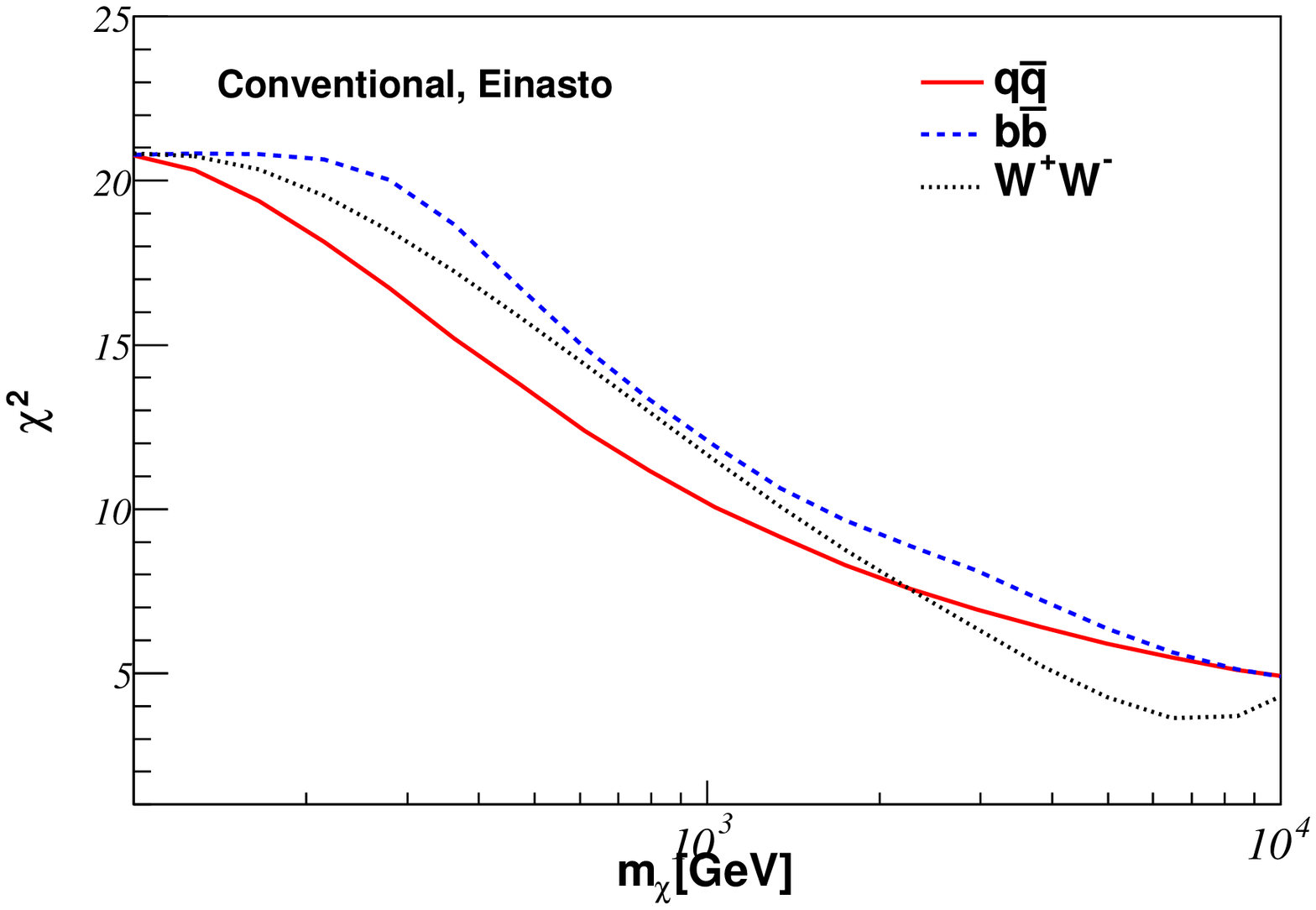}
\includegraphics[width=0.45\textwidth]{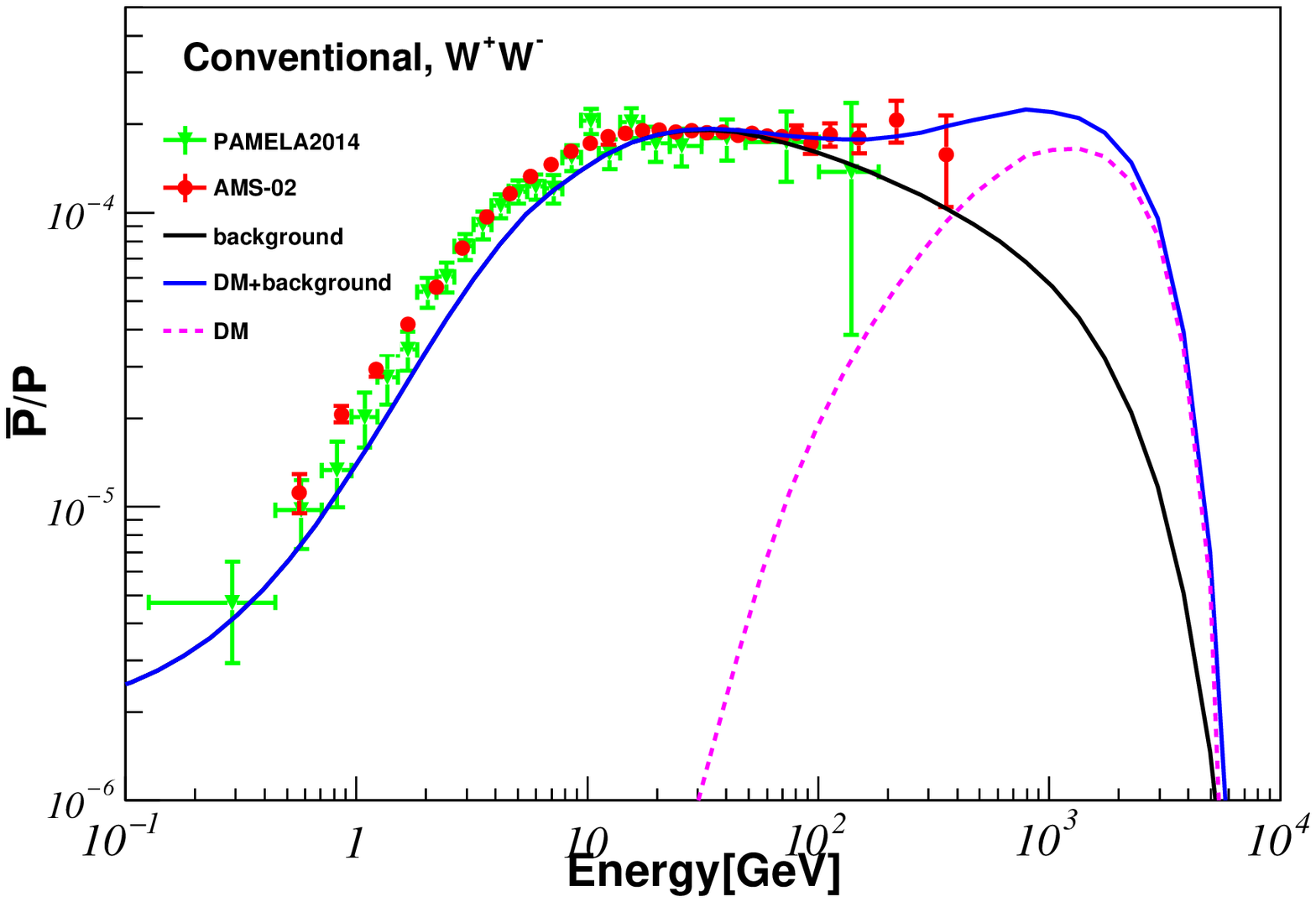}
\\
\includegraphics[width=0.45\textwidth]{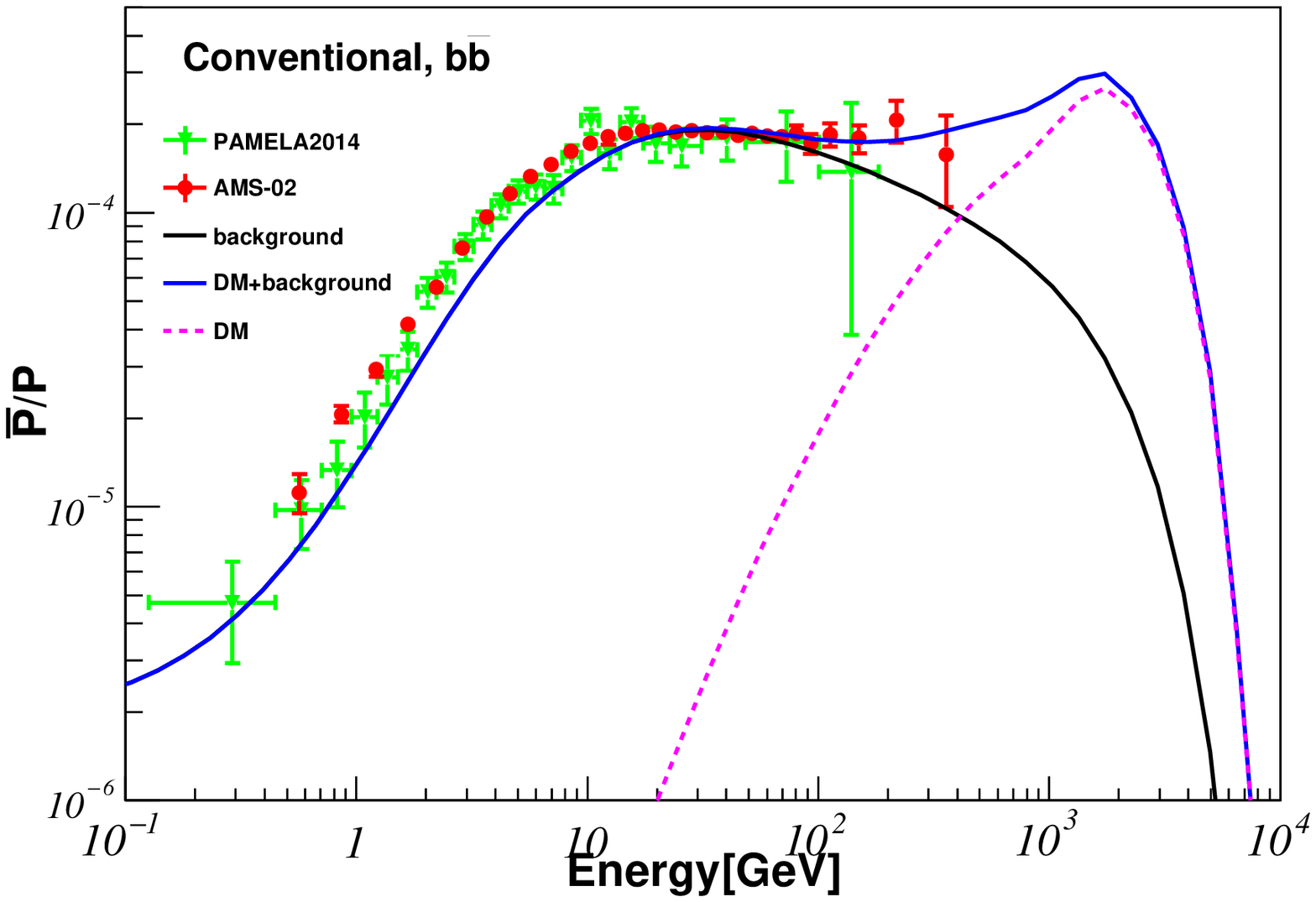}
\includegraphics[width=0.45\textwidth]{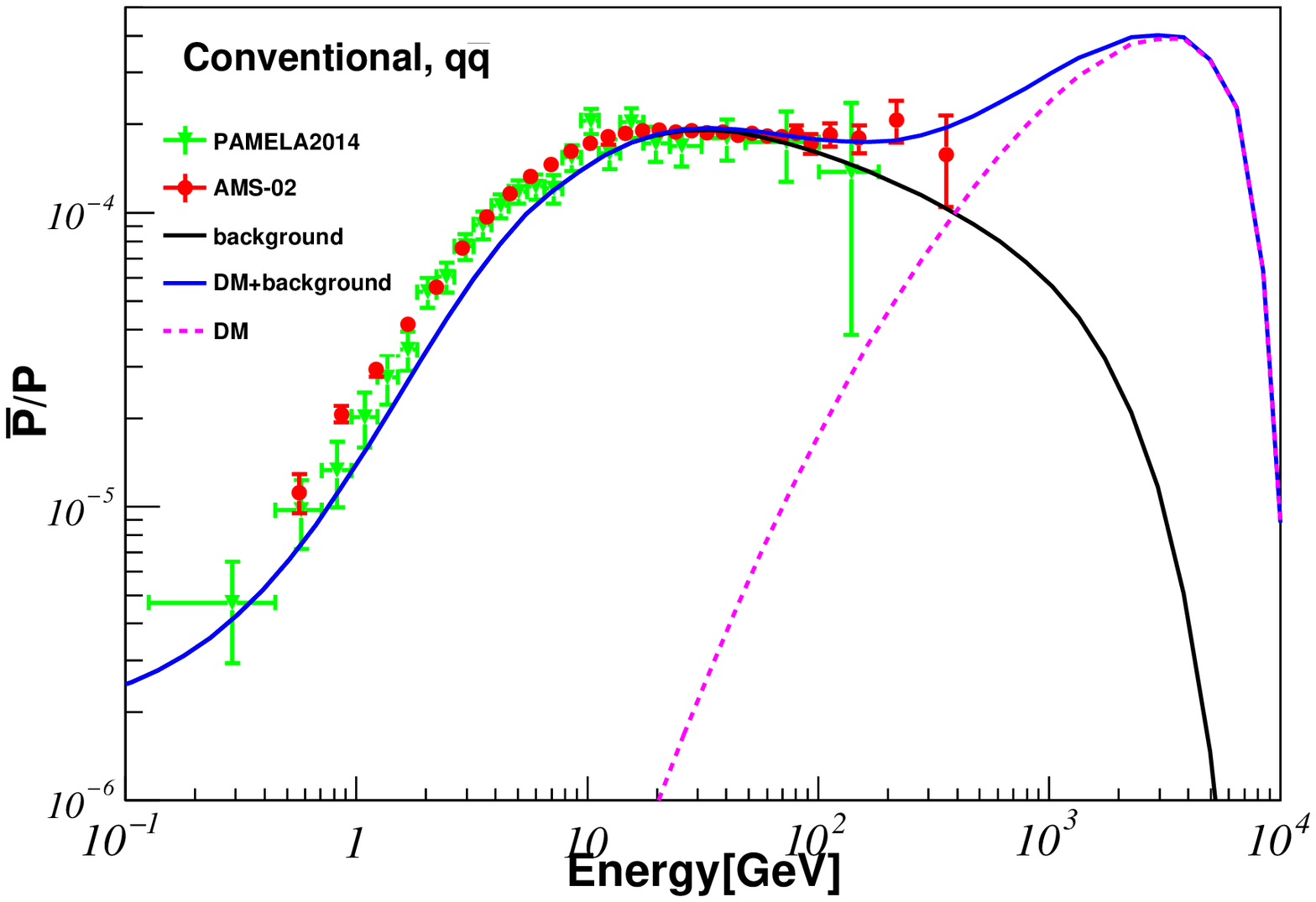}
\caption{
(Upper left)
values of $\chi^{2}_{\text{min}}$ as a function of 
DM particle mass $m_{\chi}$ from 
a fit to the AMS-02 $\bar p/p$ data ( with kinetic energy above 20 GeV ) in 
the ``conventional'' propagation model
~\cite{astro-ph/0101068,astro-ph/0510335}  with  
the DM profile  fixed to  Einasto~\cite{Einasto:2009zd%
}.
Three annihilation channels  $b\bar b$, $q\bar q$ and 
$W^{+}W^{-}$ are considered.
(Upper right)
predicted $\bar p/p$ ratio in the case of 
background (``conventional'' model)
plus a DM contribution with 
$m_{\chi}=6.5$~TeV,
$\langle \sigma v\rangle=1.9\times 10^{-24}\text{cm}^{3}\text{s}^{-1}$,
and annihilation final states $W^{+}W^{-}$.
The flux ratio of antiproton from DM to the proton from the  background  
$\bar p_{\text{DM}}/p_{\text{BG}}$ is shown as the dashed line.
The data  from AMS-02~\cite{Ting:AMS}
and PAMELA~\cite{Adriani:2014pza} are also shown.
(Lower left)
the same as the upper right, but for the $b\bar b$ channel
with
$m_{\chi}=10.9$~TeV and 
$\langle \sigma v\rangle=3.4\times 10^{-24}~\text{cm}^{3}\text{s}^{-1}$.
(Lower right)
the same as the upper right, but for the $q\bar q$ channel with
$m_{\chi}=10.9$~TeV and 
$\langle \sigma v\rangle=3.3\times 10^{-24}~\text{cm}^{3}\text{s}^{-1}$.
}
\label{fig:chisq}
\end{center}
\end{figure}

As can be seen from \fig{fig:pbar}, 
compared with the AMS-02 data
the GALPROP  DR models predict fewer antiprotons at
low ($\lesssim 10$ GeV) and very high ($\gtrsim 100$ GeV ) energies.
Without a robust estimation of the theoretical uncertainties,
it is premature to claim any excesses in the $\bar p/p$ data.
Nevertheless, 
we consider what would be the implications for DM  
if such a trend in the observations is confirmed by  future analyses. 
The low energy data would allow for a non-vanishing DM annihilation cross section.
For instance, 
in  the ``conventional'' propagation model, 
for $m_{\chi}=$10.1, 35.0 and 75.8~GeV, 
the best-fit  values are
$\langle \sigma v\rangle =3.6\times 10^{-27}$,
$1.14\times 10^{-26}$, and 
$2.79\times 10^{-26}\text{ cm}^{3}\text{s}^{-1}$, 
respectively, 
if the DM profile is Einasto, 
and the DM particles annihilate dominantly into $\bar b b$ final states.
If both $m_{\chi}$ and $\langle \sigma v\rangle$ are allowed to vary freely,
the best-fit  DM particle masses and annihilation cross sections  are 
$$
m_{\chi}=58.5\ (35.0)~\text{GeV and } 
\langle \sigma v\rangle=2.16\ (0.86)\times 10^{-26}\text{ cm}^{3}\text{s}^{-1}
$$
for DM annihilating into $\bar b b\ (\bar q q)$ final states.
In \fig{fig:low-energy-fit}, 
we show the calculated spectra of $\bar p/p$ flux ratio from 
the best-fit DM particle masses and cross sections.
The figure shows that the low energy $\bar p/p$ data are well reproduced
by including such a DM contribution,
except for the data point with kinetic energy below 1~GeV.

As shown in \fig{fig:pbar}, 
the spectrum of the  AMS-02 $\bar p/p$ ratio tends to be flat toward 
high energies above $\sim100$ GeV.
This trend, if confirmed by the future AMS-02 data,
is not expected from the secondary production of antiprotons,
and 
raises the interesting question whether 
this would leave some room for a heavy DM contribution,
similar to the case of the AMS-02 positron fraction
\cite{Accardo:2014lma, %
Kopp:2013eka,%
Bergstrom:2013jra, %
Jin:2013nta,Liu:2013vha}.
To explore this possibility,
we perform an other  fit using
the $\bar p/p$ ratio data above 20~GeV (15 data points in total)
in order to avoid the theoretical uncertainties in the low energy region.
The obtained  
$\chi^{2}_{\text{min}}$ as a function of $m_{\chi}$ for
the $b\bar b$, $q\bar q$ and $W^{+}W^{-}$ final states in
the  ``conventional'' propagation model with Einasto DM profile
are shown in \fig{fig:chisq}.
One can see that
for the three final states
the values of $\chi^{2}_{\text{min}}$  decrease almost monotonically 
from $\sim21$ to $\sim5$ with 
an increasing DM particles mass from $100$~GeV to $10$~TeV,
but
the $\chi^{2}$-curves become gradually  flat toward high DM masses.
Only for the $W^{+}W^{-}$ channel, there exists a shallow  local minimal at
around 6.5~TeV with low statistical significance. 
From the $\chi^{2}$-curves, 
one can see that the DM particles mass is restricted to be
above $\sim2$~TeV at $2\sigma$. 
For an illustration purpose, 
we  show in \fig{fig:chisq} 
the predictions for the  $\bar p/p$ ratio in  
the ``conventional''  background model  with a DM contribution. 
The DM particles masses and annihilation cross sections  chosen to be 
$m_{\chi}=6.5$~TeV,
$\langle \sigma v\rangle=1.9\times 10^{-24}\text{cm}^{3}\text{s}^{-1}$
for $W^{+}W^{-}$,
$m_{\chi}=10.9$~TeV,
$\langle \sigma v\rangle=3.4\times 10^{-24}~\text{cm}^{3}\text{s}^{-1}$
for $b\bar b$ channel, 
and 
$m_{\chi}=10.9$~TeV and 
$\langle \sigma v\rangle=3.3\times 10^{-24}~\text{cm}^{3}\text{s}^{-1}$
for $q\bar q$ channel.
Note that these values are not from the best-fit values.
We conclude that 
introducing a DM contribution can  improve the 
agreement with the AMS-02 $\bar p/p$ data with kinetic energy above 100~GeV,
but the statistics is not high enough to
determine  the DM properties such as its mass and interaction strength.
If the DM particle mass is indeed at $\mathcal{O}(10)$ TeV scale, 
next generation precision cosmic-ray detection experiments are 
needed. 
As can be see in \fig{fig:pbar}, 
the possible ``excess'' is located at the kinetic energy range $100-450$~GeV where
the secondary backgrounds from the four propagation models are similar.
However, beyond $\sim 450$ GeV, the  $\bar p/p$ from the ``conventional'' model 
drops quicker than that in the other propagation models. 
The future high energy antiproton data will be very important not only in 
probing DM but also in constraining the background models.

In summary,
we have explored the significance of the first  AMS-02 $\bar p/p$ data on 
constraining the annihilation cross sections of the DM particles in 
various propagation models and DM profiles. 
Four representative background models have been considered with 
four different DM profiles.
We have  derived the upper limits using the GALPROP code
and 
shown that 
in the ``conventional '' propagation model with  Einasto DM profile, 
the constraints can be more stringent than that derived from
the Ferm-LAT gamma-ray data on 
the dwarf spheroidal satellite galaxies.
Making use of  the typical 
minimal, median and maximal models obtained from 
a previous global fit, we have shown that
the uncertainties on the upper limits is around a factor of five.
The future more precise AMS-02 data can help to reduce the uncertainties 
in the derived upper limits. 

{\it Note  added:}
As we were completing this study, 
Ref.
\cite{Giesen:2015ufa} 
appeared on the  arXiv, 
which addresses some of the same problems as discussed here.
Although the conclusions are similar, 
the analysis in this  work  is based on the fully numerical GALPROP code, 
while that in \cite{Giesen:2015ufa}  is based on the two-zone diffusion model
with (semi)-analytical approach. The two methods are quite complementary to 
each other. 
Compared with Ref~\cite{Giesen:2015ufa}, 
the upper limits obtained in this work are  weaker at DM particle mass 
below $\sim 100$ GeV, but stronger for heavy DM particles above $\sim 500$ GeV
for typical DM particles annihilating  into $\bar b b$ final states.
Similar discussions on the DM matter contributions can be found in 
Refs.~\cite{
Ibe:2015tma,
Chen:2015cqa%
}.

\subsection*{Acknowledgments}
YLW is grateful to  S. Ting for warm hospitality and 
insightful discussions during his visit to the AMS-02 POCC at CERN.
We thank P. Zuccon, A. Kounine, A. Oliva and S. Haino for 
helpful discussions on the details of the AMS-02 detector. 
This work is supported in part by
the National Basic Research Program of China (973 Program) under Grants 
No. 2010CB833000;
the National Nature Science Foundation of China (NSFC) under Grants 
No. 10905084,
No. 11335012 and
No. 11475237;
The numerical calculations were done  using
the HPC Cluster of SKLTP/ITP-CAS.

\bibliography{amsfit_inspire,misc}
\bibliographystyle{JHEP}

\end{document}